\documentclass[aps,prl,twocolumn,superscriptaddress,showpacs,floatfix]{revtex4-2}
\usepackage[dvipdfmx]{graphicx,color,hyperref}

\begin{document}
\title{Dispersive measurement of spin shot noise in a Bose--Einstein condensate}
\author{Kosuke Shibata}
\email{shibata@qo.phys.gakushuin.ac.jp}
\affiliation{Department of Physics, Gakushuin University, Tokyo 171-8588, Japan}
\affiliation{Department of Physics, Graduate School of Science, Kyoto University, Kyoto 606-8502, Japan}
\author{Naota Sekiguchi}
\affiliation{Department of Electrical and Electronic Engineering, Institute of Science Tokyo, Tokyo 152-8550, Japan}
\author{Junnosuke Takai}
\affiliation{National Metrology Institute of Japan, National Institute of Advanced Industrial Science and Technology, Ibaraki 305-8563, Japan }
\author{Takuya Hirano}
\affiliation{Department of Physics, Gakushuin University, Tokyo 171-8588, Japan}
\date{\today}

\begin{abstract}
We report dispersive spin shot noise measurement of a Bose--Einstein condensate (BEC).
While dispersive probing has been used for quantum spin noise measurement of thermal and cold gases for decades, 
confirmative measurement of spin shot noise, i.e.,\ the linear dependence of the spin variance on the number of atoms in a BEC has been lacking. 
Here, we demonstrate precise spin noise measurement of a BEC of rubidium atoms at the spin shot noise level 
by polarization rotation using a two-color probe at optimal detunings, with power balance stabilization to suppress probe-induced excess spin noise.
This work opens the possibility for the unexplored study of quantum spin fluctuations in multi-component or spinor BECs
and offers an approach to improve spin measurement precision, which is relevant to atomic spin-based sensors.
\end{abstract}

\maketitle

Spin, an inherently quantum mechanical quantity, now plays an important role in cutting-edge applications including quantum information processing and quantum computation \cite{Watson2018, Arute2019, Xue2021, Barnes2022}. 
It can also serve as a high-precision quantum sensor \cite{Degen2017}.
For example, atomic vapor is one of the most sensitive magnetometers with a sensitivity down to sub fT$/\sqrt{\mathrm{Hz}}$ \cite{Kominis2003,Dang2010}. 
A cold atomic gas, especially a Bose--Einstein condensate (BEC), is a good candidate for a spatially resolved magnetometer \cite{Wildermuth2005, Vengalattore2007, Yang2017, Sekiguchi2021}. 
It has been pointed out that a tightly confined single-domain BEC can achieve an energy resolution, which is a measure of the magnetic field sensitivity obtainable with a given measurement volume, of better than $\hbar$ \cite{Alvarez2022},
exceeding the limit for superconducting quantum interference devices \cite{Koch1980, Koch1981}.

Precise spin measurement is a key ingredient for the aforementioned applications.
A challenge in spin measurement is to achieve sensitivity at the standard quantum limit (SQL), 
the minimum sensitivity attainable with quantum-mechanically uncorrelated probes (atoms in the context here) \cite{Giovannetti2006, Boixo2007, Pezze2018}.
As the SQL is proportional to $1/\sqrt{N}$, where $N$ is the number of probes \cite{Braunstein1994}, it can be easily hindered by technical noise, particularly for a large system.
Nevertheless, the spin shot noise for atomic gases has been measured and 
even spin squeezing induced by measurement has been reported for hot \cite{Vasilakis2015, Bao2020, Kong2020} and cold \cite{Appel2009, 
Takano2009, Sewell2012, Colangelo2017, Hemmer2021} gases. 
Squeezing of a pseudospin consisting of two hyperfine sublevels (atom number squeezing) of a BEC has been observed with an absorptive atom number measurement \cite{Esteve2008, Gross2010, Muessel2014}.
However, confirmative evidence of successful shot noise measurement
of real spin, i.e.,\ the linear increase of the spin variance with respect to the atom number, has not been observed for a BEC.

Spin measurement of a BEC at the SQL level will open a new avenue in the study of multi-component or spinor BECs \cite{StamperKurn2013},
in addition to the pursuit of the best measurement precision. 
The observation of quantum spin fluctuations will help to experimentally explore the fundamental aspects of BECs,
as did precise atom number measurements at the shot-noise-level \cite{Kristensen2019}, where a sudden increase of fluctuations at the critical temperature \cite{Politzer1996, Navez1997} was observed. 
Dispersive spin probing is superior to absorptive probing, which is more common in cold atom experiments,
in that it allows in situ and consecutive spin measurement,
offering the possibility to track the spin dynamics.

Polarization rotation detection, one of the most popular dispersive spin probing techniques, is also attractive in the context of quantum nondemolition (QND) spin measurement \cite{Kuzmich1998, Takahashi1999}.
Traditionally, measurement-induced spin squeezing of a cold gas has been demonstrated 
with the use of a cigar-shaped gas in an elongated optical trap \cite{Appel2009,Sewell2012, Colangelo2017, Hemmer2021}, 
which is advantageous for significant squeezing \cite{Hammerer2004, Baragiola2014}.
For a BEC, a large atom-field interaction may be achieved without a cigar-shaped configuration
due to its high atomic density.
Therefore, dispersive spin BEC measurement may enable quantum-enhanced sensing in a small volume.
Improving the sensitivity of spin measurement will lead to breaking of the conventional magnetic field sensitivity limit in the micrometer scale and shed light on the open question of the energy resolution limit \cite{Mitchell2020}.

We report the measurement of spin noise of a BEC by polarization rotation. 
A two-color probe with stabilized power balance was used to suppress the nonlinear ac Stark shift, 
which can be a dominant technical noise source in spin shot noise measurements of BECs with a relatively small number of atoms.
The frequencies of two probe fields were optimized to reduce atom loss due to light assisted collisions, 
which tends to be significant in probing BECs due to their high atomic density. 
A CCD camera was used in the spin noise measurements to observe the polarization rotation
instead of a balanced photodiode detector, which has often been used to achieve shot-noise-level sensitivity \cite{Windpassinger2009, Ciurana2016}.
The camera naturally offers the ability to perform spatially resolved measurements
and will enable the exploration of the spatial properties of the spin noise distribution, which have not yet been studied experimentally.
The camera also offers the advantage of good effective spatial mode matching between atoms and probes when combined with post selection of the region of interest.
We observed a linear increase in the variance of the polarization rotation angle with respect to atom number,
indicating successful observation of spin shot noise
and realization of precise spin noise measurement at the SQL.

\begin{figure}[t]
	\centering
	\includegraphics[clip, scale=0.32]{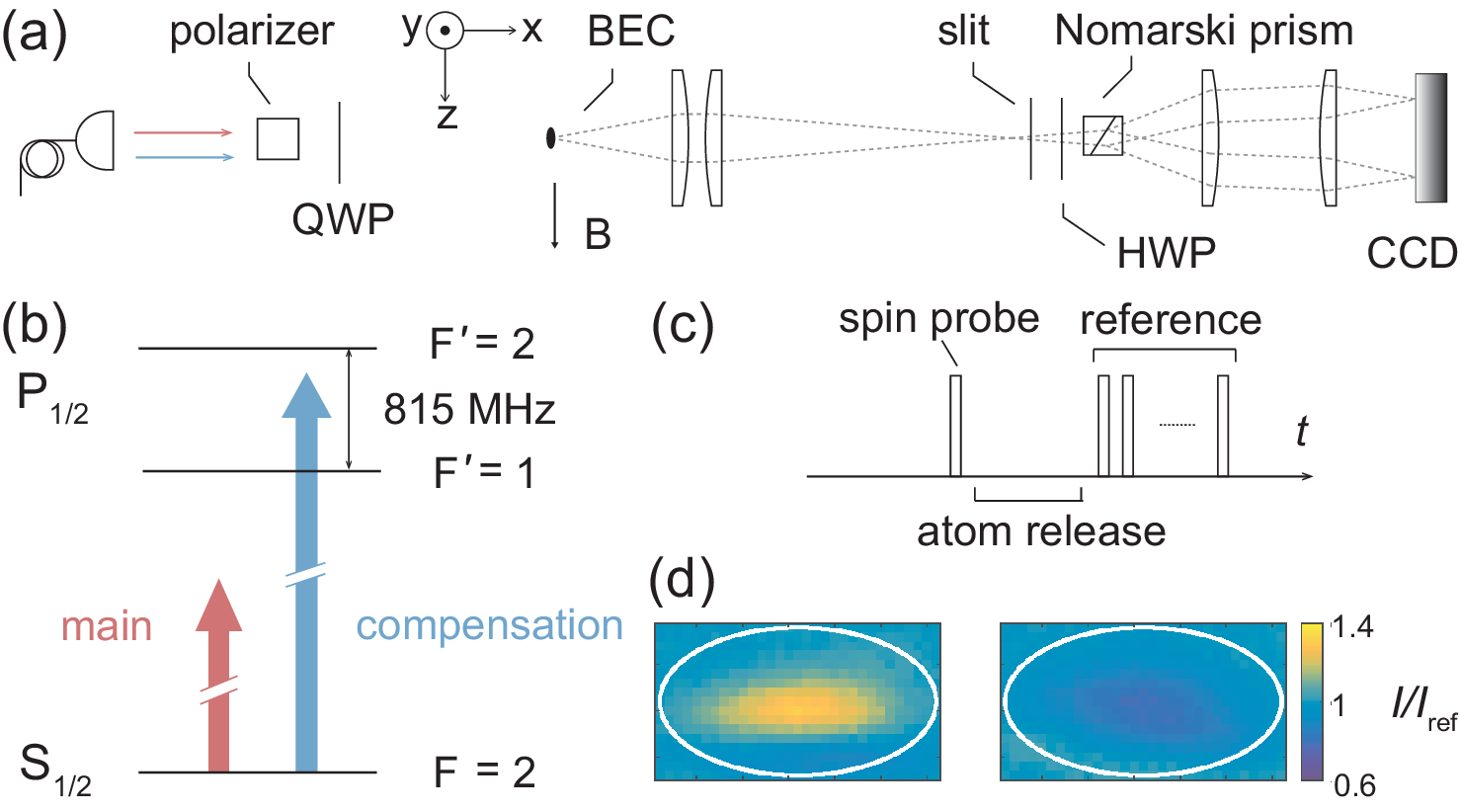}
	\caption{(Color online) Experimental setup. (a) Spin detection system. The polarization rotation for the two-color probe linearly polarized along the $z$ axis is detected by the CCD camera.  (b) Relevant energy levels of $^{87}$Rb and probe beam frequencies. (c) Time sequence of the probing pulses. (d) Images of orthogonal polarization components for a BEC with a spin direction along the $x$ axis. The measured intensity $I$ normalized against the reference intensity $I_{\mathrm{ref}}$ is shown. The white ellipse represents the boundary of the region of interest. The lengths of the major and minor axes of the ellipse are 17 $\mu$m and 9 $\mu$m, respectively.
	}
	\label{fig: setup}
\end{figure}

We measured the spin noise for a BEC in a crossed optical dipole trap.
Atoms were loaded into the trap from a magnetic trap and evaporatively cooled to produce 
an almost pure BEC in the $|F=2,m_z =+2\rangle$ state
of up to $3 \times 10^5$ atoms, where the quantized axis $z$ is defined by the direction of the bias magnetic field $B$ (see Fig.~\ref{fig: setup}(a)). 
The typical radial and axial trap frequencies of the optical trap after the final stage of the evaporation were 113(2) and 34(2) Hz, respectively. 
The bias magnetic field was set to 29.5 $\mu$T during the spin noise measurements.
The bias field was sufficiently larger than the drift of the environmental magnetic field along the $x$ axis
(typically 20 nT over a night), suppressing the variation in the spin direction due to the fluctuation of $B_x$ 
during data acquisition.
We also performed spin cleaning with a microwave transition and hyperfine-state-selective blast light to ensure an initial spin polarization.

We estimated the collective spin component, $F_x$, from the polarization rotation of the probe light. 
The probe consists of two (main and compensation) beams near the D1 line at 795 nm \cite{Takai2023}, as described below. 
The main beam frequency was stabilized using the modulation transfer spectroscopy signal of a magnetically shielded rubidium cell
with detuning from atomic resonance produced by acousto-optic modulators (AOMs).
The compensation beam was frequency-offset locked to the main beam for flexible detuning. 
The two beams were combined and coupled into a common single mode optical fiber for spatial beam mode matching. They also passed through a common AOM for pulsing beams.
The probe passed through a polarizer and a quarter waveplate (QWP) for compensation of polarization ellipticity before the atom cell.
The polarization rotation angle was measured by a polarimeter consisting of a half waveplate (HWP), a Nomarski prism, and a CCD camera capable of kinetic imaging (PIXIS1024BR, Princeton Instruments).
The polarization plane was rotated by an angle $\theta = g F_x$, where $g$ is the coupling strength determined by the atomic optical density and the probe frequency, due to the paramagnetic Faraday rotation effect. 
The variance of the polarization angle is given by
\begin{equation}\label{Var}
\mathrm{Var}(\theta) = a N + b N^2 + \mathrm{Var}(\theta)_0,
\end{equation}
where 
$\mathrm{Var}(\theta)_0 $ is the variance in the absence of atoms, including the variance due to photon shot noise, given by $1/(4N_{\mathrm{p}})$ with $N_{\mathrm{p}}$ being the number of probe photons,
and the readout noise.
The first and second terms on the right-hand side of Eq.~(\ref{Var}) denote the spin shot noise and the technical noise, respectively, with different dependences on the atom number $N$.
Because we probed atoms in the coherent spin state, 
the spin shot noise term should be $g^2 |\langle F_{\mathrm{eff}} \rangle|/2$, 
where $F_{\mathrm{eff}} = \chi F_z$ represents the effective spin size
with the coefficient $\chi$ determined by the geometry of the atoms and the probe \cite{Baragiola2014}.

Probe-induced spin change is one of the dominant types of technical noise sources in spin measurements.
This is particularly true in our spin shot noise measurements of a BEC.
Measuring spin shot noise for a small number of atoms 
requires a large probe photon number to reduce uncertainties due to photon shot noise 
in polarization rotation measurements, thus increasing probe-induced spin changes.
For a quantitative discussion, we introduce the ratio of the spin shot noise of the coherent spin state (without consideration of the detection geometry \cite{Baragiola2014}) to the photon shot noise resolution in the polarization rotation measurement \cite{Deutsch2010}, 
$\xi = 4g^2 N N_{\mathrm{p}}$. 
A larger $\xi$ is better for spin shot noise measurements. 
As the coupling strength $g$ scales to $(A\delta)^{-1}$, where $A$ is the measurement cross section
and $\delta$ is the detuning from the resonance (a two-level model is assumed here  for simplicity), $\xi$ can be written as 
\begin{equation}
\xi = 4\tilde{g}^2 \frac{N}{A}\frac{N_{\mathrm{p}}}{A\delta^2},
\end{equation}
where $\tilde{g}$ is the normalized coupling strength, independent of $A$ and $\delta$.
Spin shot noise measurements of an atom cloud of small column atom density ($N/A$) require a large $N_{\mathrm{p}}/(A\delta^2)$,
which causes large photo-induced atom loss, photon scattering, and light shift.
Spin-dependent light shift due to vector and tensor atom-field interaction
is harmful for precise spin measurement, as explained below.
The column atom density is not high in our measurement configuration
where a BEC is probed along its radial direction, 
while the configuration is compatible with spatially resolved magnetometry \cite{Sekiguchi2021}.
Therefore, we make efforts to reduce the spin-dependent light shift.

The vector light shift or ``fictitious magnetic field'' 
acts in the same manner as a magnetic field along the probe propagation axis and causes instantaneous spin rotation during the probe pulse.
It increases technical noise in a spin measurement when combined with shot-to-shot fluctuations of the probe intensity and magnetic field.
We suppress the vector light shift by reducing the ellipticity of the probe.
Such a reduction is often optimized by spectroscopy with atoms themselves \cite{Steffen2013, Wood2016}.
Spectroscopy is, however, less sensitive here because the fictitious field is orthogonal to the bias magnetic field.
We instead directly measured the spin change due to 
three probe pulses of 1 $\mu$s with an interval of the Larmor precession
using spin-resolved absorption imaging \cite{Shibata2021}, from which we measured the magnetization along the $z$ axis.
A $\pi$/2 pulse was applied after probe irradiation to maximize the sensitivity to the spin rotation due to the vector shift around the total magnetic field almost along the $z$ axis.
We adjusted the QWP angle to minimize the magnetization change before obtaining the experimental data presented below.

The linearly polarized probe caused spin evolution by nonlinear light shift (proportional to the square of the magnetic quantum number) 
due to the tensor term of the atom--field interaction.
Several techniques have been developed to reduce nonlinear spin evolution, such as
magic angle probing of rotating spin \cite{Smith2004}, alternating polarization probe
for a $^{87}$Rb thermal gas in the $F = 1$ state \cite{Koschorreck2010}, 
and two-color probing for Cs of $F = 4$ \cite{Hemmer2021}.

If the probe polarization vector perfectly coincided with the atomic spin polarization,
it would only induce a phase shift on a single magnetic sublevel ($|F=2,m_z =+2\rangle$ in our case) and cause no nonlinear spin evolution.
In practice however it is difficult to sufficiently eliminate the nonlinear spin evolution by this polarization matching strategy.
For example, fluctuation of the direction of the bias magnetic field defining the spin quantization axis and slow change of light polarization increases the polarization rotation variance.
Inhomogeneity of the light polarization over the atom cloud may
limit the suppression of the nonlinear evolution.

\begin{figure}[t]
	\centering
	\includegraphics[clip, scale=0.42]{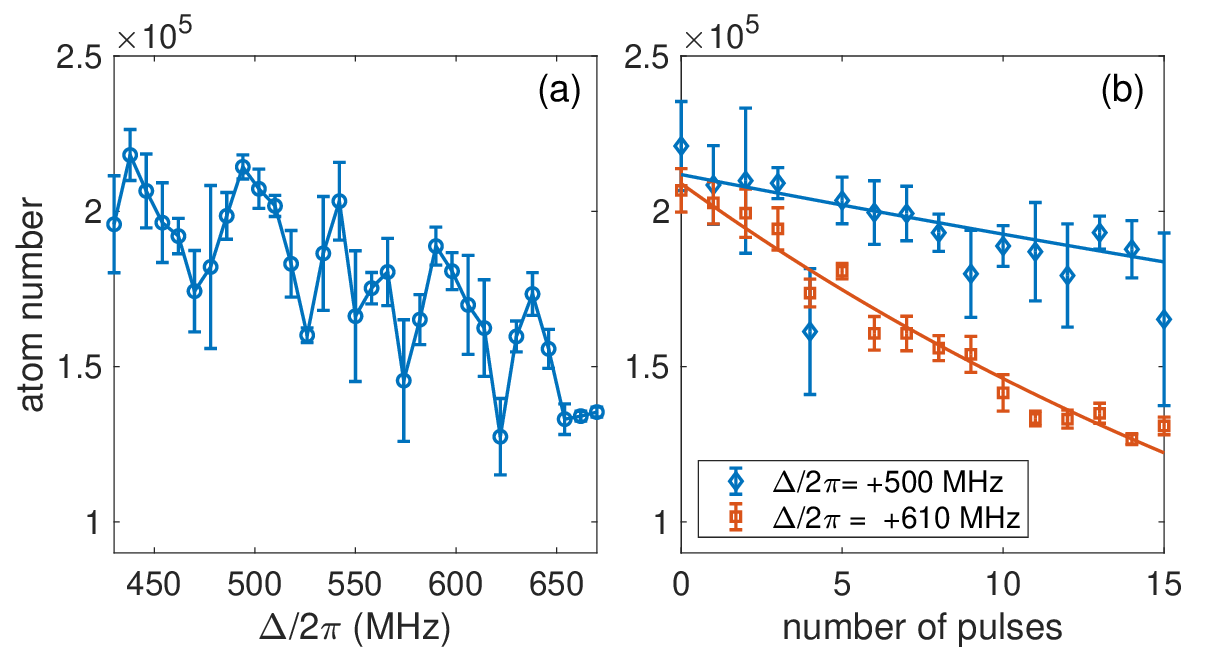}
	\caption{(Color online) Loss by the compensation beam. (a) Loss dependence on the detuning. We applied five pulses with $1$ $\mu$s width and peak power of 180 $\mu$W. (b) Loss rate measurement results. The blue diamonds and red squares represent the remaining atom number after irradiating by compensation beam pulses of $\Delta/(2\pi) = +500$ MHz and $+610$ MHz, respectively. The pulse width is 1 $\mu$s. The peak power is 180 $\mu$W. 
The solid lines are exponential fits to the data.	
	}
	\label{fig: loss rate}
\end{figure}

We adopted a two-color probe to sufficiently suppress nonlinear spin evolution.
We combined a beam with red-detuning from the $F=2$--$F'=1$ transition (main beam) and another beam with its frequency between the $F=2$--$F'=1$ and $F=2$--$F'=2$ transitions (compensation beam) [see Fig.~\ref{fig: setup}(b)]. 
The total nonlinear shift caused by these beams can be cancelled by adjusting the beam power balance \cite{Takai2023}.
The power ratio of the main and compensation beams was stabilized 
to yield an instability of less than 0.01.

Before performing measurements with the two-color probe,
we performed loss spectroscopy with the compensation beam to find its optimal detuning. 
We found several frequencies with smaller atom loss rates, as shown in Fig.~\ref{fig: loss rate}(a).
Figure~\ref{fig: loss rate}(b) shows the loss rates at certain frequencies. 
The atoms in the $F=1$ state were removed by a beam resonant to the $F=1$--$F'=0$ D$_2$ transition after each probe pulse in these measurements. 
We observed the loss rate reduction by the $F=1$ removal, 
which is attributed to the suppression of hyperfine Raman superradiant scattering \cite{Vengalattore2007}.
The atom number after $n$ pulses was fitted by $N^{(n)} = N^{(0)}\exp(-\gamma n)$,
where $N^{(0)}$ is the initial atom number.
The loss rate, $\gamma$, by the beams of $\Delta/(2\pi) = +500$ MHz and $+610$ MHz were $0.009(2)$ and $0.036(2)$, respectively.
We set the compensation beam detuning to $+500$ MHz in the following experiments.

Although we found the optimal compensation beam frequency, 
the optimal compensation beam still induces a larger atom loss than the optimally red-detuned main beam \cite{Sekiguchi2021}.
This is probably because the light-assisted collisional loss is not sufficiently suppressed for the compensation beam, 
the frequency of which is close to the atomic resonances,
and, more importantly, is blue-detuned from the $F=2$--$F'=1$ transition
with a continuous loss spectrum \cite{Burnett1996}.
The compensation beam red-detuned from the D$_2$ line may be used for a lower loss rate.

\begin{figure*}[t]
	\centering
	\includegraphics[clip, scale=0.59]{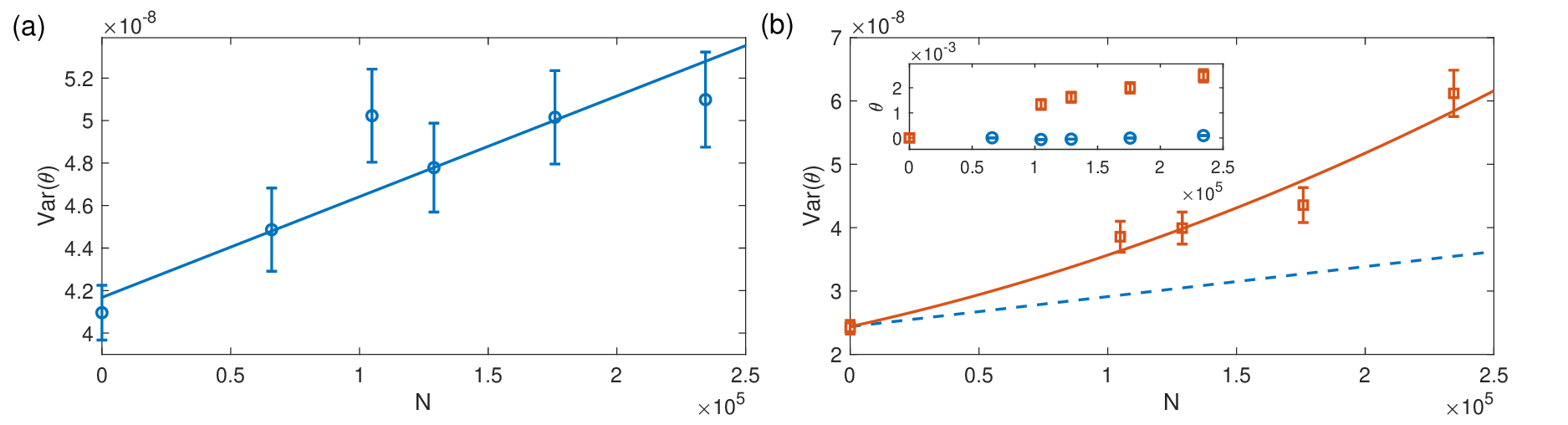}
	\caption{(Color online) Spin noise measurement. (a)  Variance of $\theta$ versus the atom number for the measurement with the two-color probe. The blue solid line is a linear fit to the measurement data (blue circles).  (b) Variance of $\theta$ without the compensation beam. The red sold line is a quadratic fit to the measurement data (red squares). The reference noise level with the same slope as (a) is also plotted (blue dashed line). The offset of the reference is adjusted to that of the red solid line. The error bars represent the statistical uncertainties of the variance. The inset shows the polarization rotation angle with (blue circles) and without (red squares) the compensation beam.}
	\label{fig: variance}
\end{figure*}

We performed repeated measurements of the polarization rotation of the probe passing the BEC in the optical trap to deduce the spin noise.
To avoid an increase in noise due to the measured slow polarization rotation angle fluctuation correlated with the room temperature change, we took reference images just after probing the atoms (see Fig.~\ref{fig: setup}(d))
and used a net polarization rotation $\theta$ due to the atoms, calculated by $\theta_{\mathrm{w/ atom}}- \theta_{\mathrm{w/o atom}}$ where $\theta_{\mathrm{w/ atom}}$ represents the estimated polarization rotation angle for a camera image with atoms 
and $\theta_{\mathrm{w/o atom}}$ is the average of the estimated angles over the reference frames, for the noise analysis.
The polarization rotation angle was estimated from the vertically and horizontally polarized probe photon numbers, $N_{\mathrm{V}}$ and $N_{\mathrm{H}}$, in the region of interest of the imaging camera shown in Fig.~\ref{fig: setup}(e) as $(N_{\mathrm{V}}-N_{\mathrm{H}})/2(N_{\mathrm{V}}+N_{\mathrm{H}}) $.

The variance of $\theta$, $\mathrm{Var}(\theta)$, was measured for a BEC of 
several groups of different mean atom numbers with the two-color probe.
The main and compensation beam detuning were $-840$ MHz and $+500$ MHz in this measurement.
The main probe beam detuning was set to reduce the light-assisted collisional atom loss \cite{Sekiguchi2021}.
Their power ratio was stabilized to $8.57:1$. 
The total photoelectron number detected by the camera in the region of interest was $7.8 \times 10^6$ per pulse.
The atom number was controlled by changing the final rf frequency for forced evaporation in the magnetic trap.
The measured $\mathrm{Var}(\theta)$ shown in Fig.~\ref{fig: variance}(a) was fitted by a linear function with a slope of $4.7(9) \times 10^{-14}$.
No significant quadratic increase in the variance was observed.
This result indicates that we measured the spin noise at the shot-noise-limited level.
It should be noted that $\mathrm{Var}(\theta)$ for each atom number group
was obtained from more than 1000 experimental runs over approximately 11 hours.
Repeated atom probing in a single experiment cycle \cite{Takano2009,Sewell2012} was not applied because we observed that a longer interval between the signal and reference frames increased the measured photon noise,
which was probably due to vibration of the imaging system. 
We nevertheless observed no significant excess spin noise, implying that the measurement scheme is robust.

We present the polarization measurement results with the main beam only to clarify the effect of the compensation beam.
To enhance the spin change due to the probe with limited available power,
we used a metapulse consisting of two pulses with the interval set to the Larmor period,
giving almost twice as many photoelectrons as the usual pulse. 
A variance due to atoms larger than that for the two-color probe was observed.
The measured variance was fitted by a quadratic function $p_2N^2+p_1N+p_0$.
The excess noise was attributed to the nonlinear polarization rotation.
The change in the rotation angle $\theta$ itself due to atoms was observed, while $\theta$ for the two-color probe remained constant irrespective of the atom number, as shown in the inset of Fig.~\ref{fig: variance}(b).
$\theta \sim 10^{-3}$ implies that atom number fluctuation or a beam power imbalance of $10 \%$ would result in an excess variance of 
$\theta$ of $10^{-8}$ rad$^2$ and prohibit shot-noise-limited spin measurements without the compensation beam.

To conclude, we performed spin noise measurements of a BEC of $^{87}$Rb atoms trapped in an optical dipole trap by polarization rotation of a two-color probe.
Optimization of the probe frequency and polarization, and beam power balance stabilization were performed to reduce probe-induced excess noises in the polarization rotation measurement.
We observed a linear increase in the polarization rotation variance with respect to the atom number,
which is characteristic of spin shot noise.
The demonstrated precise spin noise measurements
can be used to experimentally explore the fundamental quantum properties of multi-spin-component or spinor BECs.
Precise and shot-noise-limited spin measurements are also key to improving BEC magnetometry.

\begin{acknowledgements}
This work was supported by MEXT Quantum Leap Flagship Program (MEXT Q-LEAP) Grant Number JPMXS0118070326, JSPS KAKENHI Grant Number JP22K03496, and the Asahi Glass Foundation.
\end{acknowledgements}


\begin{thebibliography}{48}%
\makeatletter
\providecommand \@ifxundefined [1]{%
 \@ifx{#1\undefined}
}%
\providecommand \@ifnum [1]{%
 \ifnum #1\expandafter \@firstoftwo
 \else \expandafter \@secondoftwo
 \fi
}%
\providecommand \@ifx [1]{%
 \ifx #1\expandafter \@firstoftwo
 \else \expandafter \@secondoftwo
 \fi
}%
\providecommand \natexlab [1]{#1}%
\providecommand \enquote  [1]{``#1''}%
\providecommand \bibnamefont  [1]{#1}%
\providecommand \bibfnamefont [1]{#1}%
\providecommand \citenamefont [1]{#1}%
\providecommand \href@noop [0]{\@secondoftwo}%
\providecommand \href [0]{\begingroup \@sanitize@url \@href}%
\providecommand \@href[1]{\@@startlink{#1}\@@href}%
\providecommand \@@href[1]{\endgroup#1\@@endlink}%
\providecommand \@sanitize@url [0]{\catcode `\\12\catcode `\$12\catcode
  `\&12\catcode `\#12\catcode `\^12\catcode `\_12\catcode `\%12\relax}%
\providecommand \@@startlink[1]{}%
\providecommand \@@endlink[0]{}%
\providecommand \url  [0]{\begingroup\@sanitize@url \@url }%
\providecommand \@url [1]{\endgroup\@href {#1}{\urlprefix }}%
\providecommand \urlprefix  [0]{URL }%
\providecommand \Eprint [0]{\href }%
\providecommand \doibase [0]{https://doi.org/}%
\providecommand \selectlanguage [0]{\@gobble}%
\providecommand \bibinfo  [0]{\@secondoftwo}%
\providecommand \bibfield  [0]{\@secondoftwo}%
\providecommand \translation [1]{[#1]}%
\providecommand \BibitemOpen [0]{}%
\providecommand \bibitemStop [0]{}%
\providecommand \bibitemNoStop [0]{.\EOS\space}%
\providecommand \EOS [0]{\spacefactor3000\relax}%
\providecommand \BibitemShut  [1]{\csname bibitem#1\endcsname}%
\let\auto@bib@innerbib\@empty
\bibitem [{\citenamefont {Watson}\ \emph {et~al.}(2018)\citenamefont {Watson},
  \citenamefont {Philips}, \citenamefont {Kawakami}, \citenamefont {Ward},
  \citenamefont {Scarlino}, \citenamefont {Veldhorst}, \citenamefont {Savage},
  \citenamefont {Lagally}, \citenamefont {Friesen}, \citenamefont
  {Coppersmith}, \citenamefont {Eriksson},\ and\ \citenamefont
  {Vandersypen}}]{Watson2018}%
  \BibitemOpen
  \bibfield  {author} {\bibinfo {author} {\bibfnamefont {T.~F.}\ \bibnamefont
  {Watson}}, \bibinfo {author} {\bibfnamefont {S.~G.~J.}\ \bibnamefont
  {Philips}}, \bibinfo {author} {\bibfnamefont {E.}~\bibnamefont {Kawakami}},
  \bibinfo {author} {\bibfnamefont {D.~R.}\ \bibnamefont {Ward}}, \bibinfo
  {author} {\bibfnamefont {P.}~\bibnamefont {Scarlino}}, \bibinfo {author}
  {\bibfnamefont {M.}~\bibnamefont {Veldhorst}}, \bibinfo {author}
  {\bibfnamefont {D.~E.}\ \bibnamefont {Savage}}, \bibinfo {author}
  {\bibfnamefont {M.~G.}\ \bibnamefont {Lagally}}, \bibinfo {author}
  {\bibfnamefont {M.}~\bibnamefont {Friesen}}, \bibinfo {author} {\bibfnamefont
  {S.~N.}\ \bibnamefont {Coppersmith}}, \bibinfo {author} {\bibfnamefont
  {M.~A.}\ \bibnamefont {Eriksson}},\ and\ \bibinfo {author} {\bibfnamefont
  {L.~M.~K.}\ \bibnamefont {Vandersypen}},\ }\href
  {https://doi.org/10.1038/nature25766} {\bibfield  {journal} {\bibinfo
  {journal} {Nature}\ }\textbf {\bibinfo {volume} {555}},\ \bibinfo {pages}
  {633} (\bibinfo {year} {2018})}\BibitemShut {NoStop}%
\bibitem [{\citenamefont {Arute}\ \emph {et~al.}(2019)\citenamefont {Arute},
  \citenamefont {Arya}, \citenamefont {Babbush}, \citenamefont {Bacon},
  \citenamefont {Bardin}, \citenamefont {Barends}, \citenamefont {Biswas},
  \citenamefont {Boixo}, \citenamefont {Brandao}, \citenamefont {Buell},
  \citenamefont {Burkett}, \citenamefont {Chen}, \citenamefont {Chen},
  \citenamefont {Chiaro}, \citenamefont {Collins}, \citenamefont {Courtney},
  \citenamefont {Dunsworth}, \citenamefont {Farhi}, \citenamefont {Foxen},
  \citenamefont {Fowler}, \citenamefont {Gidney}, \citenamefont {Giustina},
  \citenamefont {Graff}, \citenamefont {Guerin}, \citenamefont {Habegger},
  \citenamefont {Harrigan}, \citenamefont {Hartmann}, \citenamefont {Ho},
  \citenamefont {Hoffmann}, \citenamefont {Huang}, \citenamefont {Humble},
  \citenamefont {Isakov}, \citenamefont {Jeffrey}, \citenamefont {Jiang},
  \citenamefont {Kafri}, \citenamefont {Kechedzhi}, \citenamefont {Kelly},
  \citenamefont {Klimov}, \citenamefont {Knysh}, \citenamefont {Korotkov},
  \citenamefont {Kostritsa}, \citenamefont {Landhuis}, \citenamefont
  {Lindmark}, \citenamefont {Lucero}, \citenamefont {Lyakh}, \citenamefont
  {Mandr{\`a}}, \citenamefont {McClean}, \citenamefont {McEwen}, \citenamefont
  {Megrant}, \citenamefont {Mi}, \citenamefont {Michielsen}, \citenamefont
  {Mohseni}, \citenamefont {Mutus}, \citenamefont {Naaman}, \citenamefont
  {Neeley}, \citenamefont {Neill}, \citenamefont {Niu}, \citenamefont {Ostby},
  \citenamefont {Petukhov}, \citenamefont {Platt}, \citenamefont {Quintana},
  \citenamefont {Rieffel}, \citenamefont {Roushan}, \citenamefont {Rubin},
  \citenamefont {Sank}, \citenamefont {Satzinger}, \citenamefont {Smelyanskiy},
  \citenamefont {Sung}, \citenamefont {Trevithick}, \citenamefont
  {Vainsencher}, \citenamefont {Villalonga}, \citenamefont {White},
  \citenamefont {Yao}, \citenamefont {Yeh}, \citenamefont {Zalcman},
  \citenamefont {Neven},\ and\ \citenamefont {Martinis}}]{Arute2019}%
  \BibitemOpen
  \bibfield  {author} {\bibinfo {author} {\bibfnamefont {F.}~\bibnamefont
  {Arute}}, \bibinfo {author} {\bibfnamefont {K.}~\bibnamefont {Arya}},
  \bibinfo {author} {\bibfnamefont {R.}~\bibnamefont {Babbush}}, \bibinfo
  {author} {\bibfnamefont {D.}~\bibnamefont {Bacon}}, \bibinfo {author}
  {\bibfnamefont {J.~C.}\ \bibnamefont {Bardin}}, \bibinfo {author}
  {\bibfnamefont {R.}~\bibnamefont {Barends}}, \bibinfo {author} {\bibfnamefont
  {R.}~\bibnamefont {Biswas}}, \bibinfo {author} {\bibfnamefont
  {S.}~\bibnamefont {Boixo}}, \bibinfo {author} {\bibfnamefont {F.~G. S.~L.}\
  \bibnamefont {Brandao}}, \bibinfo {author} {\bibfnamefont {D.~A.}\
  \bibnamefont {Buell}}, \bibinfo {author} {\bibfnamefont {B.}~\bibnamefont
  {Burkett}}, \bibinfo {author} {\bibfnamefont {Y.}~\bibnamefont {Chen}},
  \bibinfo {author} {\bibfnamefont {Z.}~\bibnamefont {Chen}}, \bibinfo {author}
  {\bibfnamefont {B.}~\bibnamefont {Chiaro}}, \bibinfo {author} {\bibfnamefont
  {R.}~\bibnamefont {Collins}}, \bibinfo {author} {\bibfnamefont
  {W.}~\bibnamefont {Courtney}}, \bibinfo {author} {\bibfnamefont
  {A.}~\bibnamefont {Dunsworth}}, \bibinfo {author} {\bibfnamefont
  {E.}~\bibnamefont {Farhi}}, \bibinfo {author} {\bibfnamefont
  {B.}~\bibnamefont {Foxen}}, \bibinfo {author} {\bibfnamefont
  {A.}~\bibnamefont {Fowler}}, \bibinfo {author} {\bibfnamefont
  {C.}~\bibnamefont {Gidney}}, \bibinfo {author} {\bibfnamefont
  {M.}~\bibnamefont {Giustina}}, \bibinfo {author} {\bibfnamefont
  {R.}~\bibnamefont {Graff}}, \bibinfo {author} {\bibfnamefont
  {K.}~\bibnamefont {Guerin}}, \bibinfo {author} {\bibfnamefont
  {S.}~\bibnamefont {Habegger}}, \bibinfo {author} {\bibfnamefont {M.~P.}\
  \bibnamefont {Harrigan}}, \bibinfo {author} {\bibfnamefont {M.~J.}\
  \bibnamefont {Hartmann}}, \bibinfo {author} {\bibfnamefont {A.}~\bibnamefont
  {Ho}}, \bibinfo {author} {\bibfnamefont {M.}~\bibnamefont {Hoffmann}},
  \bibinfo {author} {\bibfnamefont {T.}~\bibnamefont {Huang}}, \bibinfo
  {author} {\bibfnamefont {T.~S.}\ \bibnamefont {Humble}}, \bibinfo {author}
  {\bibfnamefont {S.~V.}\ \bibnamefont {Isakov}}, \bibinfo {author}
  {\bibfnamefont {E.}~\bibnamefont {Jeffrey}}, \bibinfo {author} {\bibfnamefont
  {Z.}~\bibnamefont {Jiang}}, \bibinfo {author} {\bibfnamefont
  {D.}~\bibnamefont {Kafri}}, \bibinfo {author} {\bibfnamefont
  {K.}~\bibnamefont {Kechedzhi}}, \bibinfo {author} {\bibfnamefont
  {J.}~\bibnamefont {Kelly}}, \bibinfo {author} {\bibfnamefont {P.~V.}\
  \bibnamefont {Klimov}}, \bibinfo {author} {\bibfnamefont {S.}~\bibnamefont
  {Knysh}}, \bibinfo {author} {\bibfnamefont {A.}~\bibnamefont {Korotkov}},
  \bibinfo {author} {\bibfnamefont {F.}~\bibnamefont {Kostritsa}}, \bibinfo
  {author} {\bibfnamefont {D.}~\bibnamefont {Landhuis}}, \bibinfo {author}
  {\bibfnamefont {M.}~\bibnamefont {Lindmark}}, \bibinfo {author}
  {\bibfnamefont {E.}~\bibnamefont {Lucero}}, \bibinfo {author} {\bibfnamefont
  {D.}~\bibnamefont {Lyakh}}, \bibinfo {author} {\bibfnamefont
  {S.}~\bibnamefont {Mandr{\`a}}}, \bibinfo {author} {\bibfnamefont {J.~R.}\
  \bibnamefont {McClean}}, \bibinfo {author} {\bibfnamefont {M.}~\bibnamefont
  {McEwen}}, \bibinfo {author} {\bibfnamefont {A.}~\bibnamefont {Megrant}},
  \bibinfo {author} {\bibfnamefont {X.}~\bibnamefont {Mi}}, \bibinfo {author}
  {\bibfnamefont {K.}~\bibnamefont {Michielsen}}, \bibinfo {author}
  {\bibfnamefont {M.}~\bibnamefont {Mohseni}}, \bibinfo {author} {\bibfnamefont
  {J.}~\bibnamefont {Mutus}}, \bibinfo {author} {\bibfnamefont
  {O.}~\bibnamefont {Naaman}}, \bibinfo {author} {\bibfnamefont
  {M.}~\bibnamefont {Neeley}}, \bibinfo {author} {\bibfnamefont
  {C.}~\bibnamefont {Neill}}, \bibinfo {author} {\bibfnamefont {M.~Y.}\
  \bibnamefont {Niu}}, \bibinfo {author} {\bibfnamefont {E.}~\bibnamefont
  {Ostby}}, \bibinfo {author} {\bibfnamefont {A.}~\bibnamefont {Petukhov}},
  \bibinfo {author} {\bibfnamefont {J.~C.}\ \bibnamefont {Platt}}, \bibinfo
  {author} {\bibfnamefont {C.}~\bibnamefont {Quintana}}, \bibinfo {author}
  {\bibfnamefont {E.~G.}\ \bibnamefont {Rieffel}}, \bibinfo {author}
  {\bibfnamefont {P.}~\bibnamefont {Roushan}}, \bibinfo {author} {\bibfnamefont
  {N.~C.}\ \bibnamefont {Rubin}}, \bibinfo {author} {\bibfnamefont
  {D.}~\bibnamefont {Sank}}, \bibinfo {author} {\bibfnamefont {K.~J.}\
  \bibnamefont {Satzinger}}, \bibinfo {author} {\bibfnamefont {V.}~\bibnamefont
  {Smelyanskiy}}, \bibinfo {author} {\bibfnamefont {K.~J.}\ \bibnamefont
  {Sung}}, \bibinfo {author} {\bibfnamefont {M.~D.}\ \bibnamefont
  {Trevithick}}, \bibinfo {author} {\bibfnamefont {A.}~\bibnamefont
  {Vainsencher}}, \bibinfo {author} {\bibfnamefont {B.}~\bibnamefont
  {Villalonga}}, \bibinfo {author} {\bibfnamefont {T.}~\bibnamefont {White}},
  \bibinfo {author} {\bibfnamefont {Z.~J.}\ \bibnamefont {Yao}}, \bibinfo
  {author} {\bibfnamefont {P.}~\bibnamefont {Yeh}}, \bibinfo {author}
  {\bibfnamefont {A.}~\bibnamefont {Zalcman}}, \bibinfo {author} {\bibfnamefont
  {H.}~\bibnamefont {Neven}},\ and\ \bibinfo {author} {\bibfnamefont {J.~M.}\
  \bibnamefont {Martinis}},\ }\href {https://doi.org/10.1038/s41586-019-1666-5}
  {\bibfield  {journal} {\bibinfo  {journal} {Nature}\ }\textbf {\bibinfo
  {volume} {574}},\ \bibinfo {pages} {505} (\bibinfo {year}
  {2019})}\BibitemShut {NoStop}%
\bibitem [{\citenamefont {Xue}\ \emph {et~al.}(2021)\citenamefont {Xue},
  \citenamefont {Patra}, \citenamefont {van Dijk}, \citenamefont {Samkharadze},
  \citenamefont {Subramanian}, \citenamefont {Corna}, \citenamefont
  {Paquelet~Wuetz}, \citenamefont {Jeon}, \citenamefont {Sheikh}, \citenamefont
  {Juarez-Hernandez}, \citenamefont {Esparza}, \citenamefont {Rampurawala},
  \citenamefont {Carlton}, \citenamefont {Ravikumar}, \citenamefont {Nieva},
  \citenamefont {Kim}, \citenamefont {Lee}, \citenamefont {Sammak},
  \citenamefont {Scappucci}, \citenamefont {Veldhorst}, \citenamefont
  {Sebastiano}, \citenamefont {Babaie}, \citenamefont {Pellerano},
  \citenamefont {Charbon},\ and\ \citenamefont {Vandersypen}}]{Xue2021}%
  \BibitemOpen
  \bibfield  {author} {\bibinfo {author} {\bibfnamefont {X.}~\bibnamefont
  {Xue}}, \bibinfo {author} {\bibfnamefont {B.}~\bibnamefont {Patra}}, \bibinfo
  {author} {\bibfnamefont {J.~P.~G.}\ \bibnamefont {van Dijk}}, \bibinfo
  {author} {\bibfnamefont {N.}~\bibnamefont {Samkharadze}}, \bibinfo {author}
  {\bibfnamefont {S.}~\bibnamefont {Subramanian}}, \bibinfo {author}
  {\bibfnamefont {A.}~\bibnamefont {Corna}}, \bibinfo {author} {\bibfnamefont
  {B.}~\bibnamefont {Paquelet~Wuetz}}, \bibinfo {author} {\bibfnamefont
  {C.}~\bibnamefont {Jeon}}, \bibinfo {author} {\bibfnamefont {F.}~\bibnamefont
  {Sheikh}}, \bibinfo {author} {\bibfnamefont {E.}~\bibnamefont
  {Juarez-Hernandez}}, \bibinfo {author} {\bibfnamefont {B.~P.}\ \bibnamefont
  {Esparza}}, \bibinfo {author} {\bibfnamefont {H.}~\bibnamefont
  {Rampurawala}}, \bibinfo {author} {\bibfnamefont {B.}~\bibnamefont
  {Carlton}}, \bibinfo {author} {\bibfnamefont {S.}~\bibnamefont {Ravikumar}},
  \bibinfo {author} {\bibfnamefont {C.}~\bibnamefont {Nieva}}, \bibinfo
  {author} {\bibfnamefont {S.}~\bibnamefont {Kim}}, \bibinfo {author}
  {\bibfnamefont {H.-J.}\ \bibnamefont {Lee}}, \bibinfo {author} {\bibfnamefont
  {A.}~\bibnamefont {Sammak}}, \bibinfo {author} {\bibfnamefont
  {G.}~\bibnamefont {Scappucci}}, \bibinfo {author} {\bibfnamefont
  {M.}~\bibnamefont {Veldhorst}}, \bibinfo {author} {\bibfnamefont
  {F.}~\bibnamefont {Sebastiano}}, \bibinfo {author} {\bibfnamefont
  {M.}~\bibnamefont {Babaie}}, \bibinfo {author} {\bibfnamefont
  {S.}~\bibnamefont {Pellerano}}, \bibinfo {author} {\bibfnamefont
  {E.}~\bibnamefont {Charbon}},\ and\ \bibinfo {author} {\bibfnamefont
  {L.~M.~K.}\ \bibnamefont {Vandersypen}},\ }\href
  {https://doi.org/10.1038/s41586-021-03469-4} {\bibfield  {journal} {\bibinfo
  {journal} {Nature}\ }\textbf {\bibinfo {volume} {593}},\ \bibinfo {pages}
  {205} (\bibinfo {year} {2021})}\BibitemShut {NoStop}%
\bibitem [{\citenamefont {Barnes}\ \emph {et~al.}(2022)\citenamefont {Barnes},
  \citenamefont {Battaglino}, \citenamefont {Bloom}, \citenamefont {Cassella},
  \citenamefont {Coxe}, \citenamefont {Crisosto}, \citenamefont {King},
  \citenamefont {Kondov}, \citenamefont {Kotru}, \citenamefont {Larsen},
  \citenamefont {Lauigan}, \citenamefont {Lester}, \citenamefont {McDonald},
  \citenamefont {Megidish}, \citenamefont {Narayanaswami}, \citenamefont
  {Nishiguchi}, \citenamefont {Notermans}, \citenamefont {Peng}, \citenamefont
  {Ryou}, \citenamefont {Wu},\ and\ \citenamefont {Yarwood}}]{Barnes2022}%
  \BibitemOpen
  \bibfield  {author} {\bibinfo {author} {\bibfnamefont {K.}~\bibnamefont
  {Barnes}}, \bibinfo {author} {\bibfnamefont {P.}~\bibnamefont {Battaglino}},
  \bibinfo {author} {\bibfnamefont {B.~J.}\ \bibnamefont {Bloom}}, \bibinfo
  {author} {\bibfnamefont {K.}~\bibnamefont {Cassella}}, \bibinfo {author}
  {\bibfnamefont {R.}~\bibnamefont {Coxe}}, \bibinfo {author} {\bibfnamefont
  {N.}~\bibnamefont {Crisosto}}, \bibinfo {author} {\bibfnamefont {J.~P.}\
  \bibnamefont {King}}, \bibinfo {author} {\bibfnamefont {S.~S.}\ \bibnamefont
  {Kondov}}, \bibinfo {author} {\bibfnamefont {K.}~\bibnamefont {Kotru}},
  \bibinfo {author} {\bibfnamefont {S.~C.}\ \bibnamefont {Larsen}}, \bibinfo
  {author} {\bibfnamefont {J.}~\bibnamefont {Lauigan}}, \bibinfo {author}
  {\bibfnamefont {B.~J.}\ \bibnamefont {Lester}}, \bibinfo {author}
  {\bibfnamefont {M.}~\bibnamefont {McDonald}}, \bibinfo {author}
  {\bibfnamefont {E.}~\bibnamefont {Megidish}}, \bibinfo {author}
  {\bibfnamefont {S.}~\bibnamefont {Narayanaswami}}, \bibinfo {author}
  {\bibfnamefont {C.}~\bibnamefont {Nishiguchi}}, \bibinfo {author}
  {\bibfnamefont {R.}~\bibnamefont {Notermans}}, \bibinfo {author}
  {\bibfnamefont {L.~S.}\ \bibnamefont {Peng}}, \bibinfo {author}
  {\bibfnamefont {A.}~\bibnamefont {Ryou}}, \bibinfo {author} {\bibfnamefont
  {T.-Y.}\ \bibnamefont {Wu}},\ and\ \bibinfo {author} {\bibfnamefont
  {M.}~\bibnamefont {Yarwood}},\ }\href
  {https://doi.org/10.1038/s41467-022-29977-z} {\bibfield  {journal} {\bibinfo
  {journal} {Nature Communications}\ }\textbf {\bibinfo {volume} {13}},\
  \bibinfo {pages} {2779} (\bibinfo {year} {2022})}\BibitemShut {NoStop}%
\bibitem [{\citenamefont {Degen}\ \emph {et~al.}(2017)\citenamefont {Degen},
  \citenamefont {Reinhard},\ and\ \citenamefont {Cappellaro}}]{Degen2017}%
  \BibitemOpen
  \bibfield  {author} {\bibinfo {author} {\bibfnamefont {C.~L.}\ \bibnamefont
  {Degen}}, \bibinfo {author} {\bibfnamefont {F.}~\bibnamefont {Reinhard}},\
  and\ \bibinfo {author} {\bibfnamefont {P.}~\bibnamefont {Cappellaro}},\
  }\href {https://doi.org/10.1103/RevModPhys.89.035002} {\bibfield  {journal}
  {\bibinfo  {journal} {Rev. Mod. Phys.}\ }\textbf {\bibinfo {volume} {89}},\
  \bibinfo {pages} {035002} (\bibinfo {year} {2017})}\BibitemShut {NoStop}%
\bibitem [{\citenamefont {Kominis}\ \emph {et~al.}(2003)\citenamefont
  {Kominis}, \citenamefont {Kornack}, \citenamefont {Allred},\ and\
  \citenamefont {Romalis}}]{Kominis2003}%
  \BibitemOpen
  \bibfield  {author} {\bibinfo {author} {\bibfnamefont {I.~K.}\ \bibnamefont
  {Kominis}}, \bibinfo {author} {\bibfnamefont {T.~W.}\ \bibnamefont
  {Kornack}}, \bibinfo {author} {\bibfnamefont {J.~C.}\ \bibnamefont
  {Allred}},\ and\ \bibinfo {author} {\bibfnamefont {M.~V.}\ \bibnamefont
  {Romalis}},\ }\href {https://doi.org/10.1038/nature01484} {\bibfield
  {journal} {\bibinfo  {journal} {Nature}\ }\textbf {\bibinfo {volume} {422}},\
  \bibinfo {pages} {596} (\bibinfo {year} {2003})}\BibitemShut {NoStop}%
\bibitem [{\citenamefont {Dang}\ \emph {et~al.}(2010)\citenamefont {Dang},
  \citenamefont {Maloof},\ and\ \citenamefont {Romalis}}]{Dang2010}%
  \BibitemOpen
  \bibfield  {author} {\bibinfo {author} {\bibfnamefont {H.~B.}\ \bibnamefont
  {Dang}}, \bibinfo {author} {\bibfnamefont {A.~C.}\ \bibnamefont {Maloof}},\
  and\ \bibinfo {author} {\bibfnamefont {M.~V.}\ \bibnamefont {Romalis}},\
  }\href {https://doi.org/10.1063/1.3491215} {\bibfield  {journal} {\bibinfo
  {journal} {Applied Physics Letters}\ }\textbf {\bibinfo {volume} {97}},\
  \bibinfo {pages} {151110} (\bibinfo {year} {2010})}\BibitemShut {NoStop}%
\bibitem [{\citenamefont {Wildermuth}\ \emph {et~al.}(2005)\citenamefont
  {Wildermuth}, \citenamefont {Hofferberth}, \citenamefont {Lesanovsky},
  \citenamefont {Haller}, \citenamefont {Andersson}, \citenamefont {Groth},
  \citenamefont {Bar-Joseph}, \citenamefont {Kr{\"{u}}ger},\ and\ \citenamefont
  {Schmiedmayer}}]{Wildermuth2005}%
  \BibitemOpen
  \bibfield  {author} {\bibinfo {author} {\bibfnamefont {S.}~\bibnamefont
  {Wildermuth}}, \bibinfo {author} {\bibfnamefont {S.}~\bibnamefont
  {Hofferberth}}, \bibinfo {author} {\bibfnamefont {I.}~\bibnamefont
  {Lesanovsky}}, \bibinfo {author} {\bibfnamefont {E.}~\bibnamefont {Haller}},
  \bibinfo {author} {\bibfnamefont {L.~M.}\ \bibnamefont {Andersson}}, \bibinfo
  {author} {\bibfnamefont {S.}~\bibnamefont {Groth}}, \bibinfo {author}
  {\bibfnamefont {I.}~\bibnamefont {Bar-Joseph}}, \bibinfo {author}
  {\bibfnamefont {P.}~\bibnamefont {Kr{\"{u}}ger}},\ and\ \bibinfo {author}
  {\bibfnamefont {J.}~\bibnamefont {Schmiedmayer}},\ }\href
  {https://doi.org/10.1038/435440a} {\bibfield  {journal} {\bibinfo  {journal}
  {Nature}\ }\textbf {\bibinfo {volume} {435}},\ \bibinfo {pages} {440}
  (\bibinfo {year} {2005})}\BibitemShut {NoStop}%
\bibitem [{\citenamefont {Vengalattore}\ \emph {et~al.}(2007)\citenamefont
  {Vengalattore}, \citenamefont {Higbie}, \citenamefont {Leslie}, \citenamefont
  {Guzman}, \citenamefont {Sadler},\ and\ \citenamefont
  {Stamper-Kurn}}]{Vengalattore2007}%
  \BibitemOpen
  \bibfield  {author} {\bibinfo {author} {\bibfnamefont {M.}~\bibnamefont
  {Vengalattore}}, \bibinfo {author} {\bibfnamefont {J.~M.}\ \bibnamefont
  {Higbie}}, \bibinfo {author} {\bibfnamefont {S.~R.}\ \bibnamefont {Leslie}},
  \bibinfo {author} {\bibfnamefont {J.}~\bibnamefont {Guzman}}, \bibinfo
  {author} {\bibfnamefont {L.~E.}\ \bibnamefont {Sadler}},\ and\ \bibinfo
  {author} {\bibfnamefont {D.~M.}\ \bibnamefont {Stamper-Kurn}},\ }\href
  {https://doi.org/10.1103/PhysRevLett.98.200801} {\bibfield  {journal}
  {\bibinfo  {journal} {Physical Review Letters}\ }\textbf {\bibinfo {volume}
  {98}},\ \bibinfo {pages} {200801} (\bibinfo {year} {2007})}\BibitemShut
  {NoStop}%
\bibitem [{\citenamefont {Yang}\ \emph {et~al.}(2017)\citenamefont {Yang},
  \citenamefont {Koll\'ar}, \citenamefont {Taylor}, \citenamefont {Turner},\
  and\ \citenamefont {Lev}}]{Yang2017}%
  \BibitemOpen
  \bibfield  {author} {\bibinfo {author} {\bibfnamefont {F.}~\bibnamefont
  {Yang}}, \bibinfo {author} {\bibfnamefont {A.~J.}\ \bibnamefont {Koll\'ar}},
  \bibinfo {author} {\bibfnamefont {S.~F.}\ \bibnamefont {Taylor}}, \bibinfo
  {author} {\bibfnamefont {R.~W.}\ \bibnamefont {Turner}},\ and\ \bibinfo
  {author} {\bibfnamefont {B.~L.}\ \bibnamefont {Lev}},\ }\href
  {https://doi.org/10.1103/PhysRevApplied.7.034026} {\bibfield  {journal}
  {\bibinfo  {journal} {Phys. Rev. Appl.}\ }\textbf {\bibinfo {volume} {7}},\
  \bibinfo {pages} {034026} (\bibinfo {year} {2017})}\BibitemShut {NoStop}%
\bibitem [{\citenamefont {Sekiguchi}\ \emph {et~al.}(2021)\citenamefont
  {Sekiguchi}, \citenamefont {Shibata}, \citenamefont {Torii}, \citenamefont
  {Toda}, \citenamefont {Kuramoto}, \citenamefont {Fukuda},\ and\ \citenamefont
  {Hirano}}]{Sekiguchi2021}%
  \BibitemOpen
  \bibfield  {author} {\bibinfo {author} {\bibfnamefont {N.}~\bibnamefont
  {Sekiguchi}}, \bibinfo {author} {\bibfnamefont {K.}~\bibnamefont {Shibata}},
  \bibinfo {author} {\bibfnamefont {A.}~\bibnamefont {Torii}}, \bibinfo
  {author} {\bibfnamefont {H.}~\bibnamefont {Toda}}, \bibinfo {author}
  {\bibfnamefont {R.}~\bibnamefont {Kuramoto}}, \bibinfo {author}
  {\bibfnamefont {D.}~\bibnamefont {Fukuda}},\ and\ \bibinfo {author}
  {\bibfnamefont {T.}~\bibnamefont {Hirano}},\ }\href
  {https://doi.org/10.1103/PhysRevA.104.L041306} {\bibfield  {journal}
  {\bibinfo  {journal} {Phys. Rev. A}\ }\textbf {\bibinfo {volume} {104}},\
  \bibinfo {pages} {L041306} (\bibinfo {year} {2021})}\BibitemShut {NoStop}%
\bibitem [{\citenamefont {Alvarez}\ \emph {et~al.}(2022)\citenamefont
  {Alvarez}, \citenamefont {Gomez}, \citenamefont {Coop}, \citenamefont
  {Zamora-Zamora}, \citenamefont {Mazzinghi},\ and\ \citenamefont
  {Mitchell}}]{Alvarez2022}%
  \BibitemOpen
  \bibfield  {author} {\bibinfo {author} {\bibfnamefont {S.~P.}\ \bibnamefont
  {Alvarez}}, \bibinfo {author} {\bibfnamefont {P.}~\bibnamefont {Gomez}},
  \bibinfo {author} {\bibfnamefont {S.}~\bibnamefont {Coop}}, \bibinfo {author}
  {\bibfnamefont {R.}~\bibnamefont {Zamora-Zamora}}, \bibinfo {author}
  {\bibfnamefont {C.}~\bibnamefont {Mazzinghi}},\ and\ \bibinfo {author}
  {\bibfnamefont {M.~W.}\ \bibnamefont {Mitchell}},\ }\href
  {https://doi.org/10.1073/pnas.2115339119} {\bibfield  {journal} {\bibinfo
  {journal} {Proceedings of the National Academy of Sciences}\ }\textbf
  {\bibinfo {volume} {119}},\ \bibinfo {pages} {e2115339119} (\bibinfo {year}
  {2022})}\BibitemShut {NoStop}%
\bibitem [{\citenamefont {Koch}\ \emph {et~al.}(1980)\citenamefont {Koch},
  \citenamefont {Van~Harlingen},\ and\ \citenamefont {Clarke}}]{Koch1980}%
  \BibitemOpen
  \bibfield  {author} {\bibinfo {author} {\bibfnamefont {R.~H.}\ \bibnamefont
  {Koch}}, \bibinfo {author} {\bibfnamefont {D.~J.}\ \bibnamefont
  {Van~Harlingen}},\ and\ \bibinfo {author} {\bibfnamefont {J.}~\bibnamefont
  {Clarke}},\ }\href {https://doi.org/10.1103/PhysRevLett.45.2132} {\bibfield
  {journal} {\bibinfo  {journal} {Phys. Rev. Lett.}\ }\textbf {\bibinfo
  {volume} {45}},\ \bibinfo {pages} {2132} (\bibinfo {year}
  {1980})}\BibitemShut {NoStop}%
\bibitem [{\citenamefont {Koch}\ \emph {et~al.}(1981)\citenamefont {Koch},
  \citenamefont {Van~Harlingen},\ and\ \citenamefont {Clarke}}]{Koch1981}%
  \BibitemOpen
  \bibfield  {author} {\bibinfo {author} {\bibfnamefont {R.~H.}\ \bibnamefont
  {Koch}}, \bibinfo {author} {\bibfnamefont {D.~J.}\ \bibnamefont
  {Van~Harlingen}},\ and\ \bibinfo {author} {\bibfnamefont {J.}~\bibnamefont
  {Clarke}},\ }\href {https://doi.org/10.1063/1.92345} {\bibfield  {journal}
  {\bibinfo  {journal} {Applied Physics Letters}\ }\textbf {\bibinfo {volume}
  {38}},\ \bibinfo {pages} {380} (\bibinfo {year} {1981})}\BibitemShut
  {NoStop}%
\bibitem [{\citenamefont {Giovannetti}\ \emph {et~al.}(2006)\citenamefont
  {Giovannetti}, \citenamefont {Lloyd},\ and\ \citenamefont
  {Maccone}}]{Giovannetti2006}%
  \BibitemOpen
  \bibfield  {author} {\bibinfo {author} {\bibfnamefont {V.}~\bibnamefont
  {Giovannetti}}, \bibinfo {author} {\bibfnamefont {S.}~\bibnamefont {Lloyd}},\
  and\ \bibinfo {author} {\bibfnamefont {L.}~\bibnamefont {Maccone}},\ }\href
  {https://doi.org/10.1103/PhysRevLett.96.010401} {\bibfield  {journal}
  {\bibinfo  {journal} {Phys. Rev. Lett.}\ }\textbf {\bibinfo {volume} {96}},\
  \bibinfo {pages} {010401} (\bibinfo {year} {2006})}\BibitemShut {NoStop}%
\bibitem [{\citenamefont {Boixo}\ \emph {et~al.}(2007)\citenamefont {Boixo},
  \citenamefont {Flammia}, \citenamefont {Caves},\ and\ \citenamefont
  {Geremia}}]{Boixo2007}%
  \BibitemOpen
  \bibfield  {author} {\bibinfo {author} {\bibfnamefont {S.}~\bibnamefont
  {Boixo}}, \bibinfo {author} {\bibfnamefont {S.~T.}\ \bibnamefont {Flammia}},
  \bibinfo {author} {\bibfnamefont {C.~M.}\ \bibnamefont {Caves}},\ and\
  \bibinfo {author} {\bibfnamefont {J.}~\bibnamefont {Geremia}},\ }\href
  {https://doi.org/10.1103/PhysRevLett.98.090401} {\bibfield  {journal}
  {\bibinfo  {journal} {Phys. Rev. Lett.}\ }\textbf {\bibinfo {volume} {98}},\
  \bibinfo {pages} {090401} (\bibinfo {year} {2007})}\BibitemShut {NoStop}%
\bibitem [{\citenamefont {Pezz{\`{e}}}\ \emph {et~al.}(2018)\citenamefont
  {Pezz{\`{e}}}, \citenamefont {Smerzi}, \citenamefont {Oberthaler},
  \citenamefont {Schmied},\ and\ \citenamefont {Treutlein}}]{Pezze2018}%
  \BibitemOpen
  \bibfield  {author} {\bibinfo {author} {\bibfnamefont {L.}~\bibnamefont
  {Pezz{\`{e}}}}, \bibinfo {author} {\bibfnamefont {A.}~\bibnamefont {Smerzi}},
  \bibinfo {author} {\bibfnamefont {M.~K.}\ \bibnamefont {Oberthaler}},
  \bibinfo {author} {\bibfnamefont {R.}~\bibnamefont {Schmied}},\ and\ \bibinfo
  {author} {\bibfnamefont {P.}~\bibnamefont {Treutlein}},\ }\href
  {https://doi.org/10.1103/RevModPhys.90.035005} {\bibfield  {journal}
  {\bibinfo  {journal} {Reviews of Modern Physics}\ }\textbf {\bibinfo {volume}
  {90}},\ \bibinfo {pages} {035005} (\bibinfo {year} {2018})}\BibitemShut
  {NoStop}%
\bibitem [{\citenamefont {Braunstein}\ and\ \citenamefont
  {Caves}(1994)}]{Braunstein1994}%
  \BibitemOpen
  \bibfield  {author} {\bibinfo {author} {\bibfnamefont {S.~L.}\ \bibnamefont
  {Braunstein}}\ and\ \bibinfo {author} {\bibfnamefont {C.~M.}\ \bibnamefont
  {Caves}},\ }\href {https://doi.org/10.1103/PhysRevLett.72.3439} {\bibfield
  {journal} {\bibinfo  {journal} {Phys. Rev. Lett.}\ }\textbf {\bibinfo
  {volume} {72}},\ \bibinfo {pages} {3439} (\bibinfo {year}
  {1994})}\BibitemShut {NoStop}%
\bibitem [{\citenamefont {Vasilakis}\ \emph {et~al.}(2015)\citenamefont
  {Vasilakis}, \citenamefont {Shen}, \citenamefont {Jensen}, \citenamefont
  {Balabas}, \citenamefont {Salart}, \citenamefont {Chen},\ and\ \citenamefont
  {Polzik}}]{Vasilakis2015}%
  \BibitemOpen
  \bibfield  {author} {\bibinfo {author} {\bibfnamefont {G.}~\bibnamefont
  {Vasilakis}}, \bibinfo {author} {\bibfnamefont {H.}~\bibnamefont {Shen}},
  \bibinfo {author} {\bibfnamefont {K.}~\bibnamefont {Jensen}}, \bibinfo
  {author} {\bibfnamefont {M.}~\bibnamefont {Balabas}}, \bibinfo {author}
  {\bibfnamefont {D.}~\bibnamefont {Salart}}, \bibinfo {author} {\bibfnamefont
  {B.}~\bibnamefont {Chen}},\ and\ \bibinfo {author} {\bibfnamefont {E.~S.}\
  \bibnamefont {Polzik}},\ }\href {https://doi.org/10.1038/nphys3280}
  {\bibfield  {journal} {\bibinfo  {journal} {Nature Physics}\ }\textbf
  {\bibinfo {volume} {11}},\ \bibinfo {pages} {389} (\bibinfo {year}
  {2015})}\BibitemShut {NoStop}%
\bibitem [{\citenamefont {Bao}\ \emph {et~al.}(2020)\citenamefont {Bao},
  \citenamefont {Duan}, \citenamefont {Jin}, \citenamefont {Lu}, \citenamefont
  {Li}, \citenamefont {Qu}, \citenamefont {Wang}, \citenamefont {Novikova},
  \citenamefont {Mikhailov}, \citenamefont {Zhao}, \citenamefont {M{\o}lmer},
  \citenamefont {Shen},\ and\ \citenamefont {Xiao}}]{Bao2020}%
  \BibitemOpen
  \bibfield  {author} {\bibinfo {author} {\bibfnamefont {H.}~\bibnamefont
  {Bao}}, \bibinfo {author} {\bibfnamefont {J.}~\bibnamefont {Duan}}, \bibinfo
  {author} {\bibfnamefont {S.}~\bibnamefont {Jin}}, \bibinfo {author}
  {\bibfnamefont {X.}~\bibnamefont {Lu}}, \bibinfo {author} {\bibfnamefont
  {P.}~\bibnamefont {Li}}, \bibinfo {author} {\bibfnamefont {W.}~\bibnamefont
  {Qu}}, \bibinfo {author} {\bibfnamefont {M.}~\bibnamefont {Wang}}, \bibinfo
  {author} {\bibfnamefont {I.}~\bibnamefont {Novikova}}, \bibinfo {author}
  {\bibfnamefont {E.~E.}\ \bibnamefont {Mikhailov}}, \bibinfo {author}
  {\bibfnamefont {K.-F.}\ \bibnamefont {Zhao}}, \bibinfo {author}
  {\bibfnamefont {K.}~\bibnamefont {M{\o}lmer}}, \bibinfo {author}
  {\bibfnamefont {H.}~\bibnamefont {Shen}},\ and\ \bibinfo {author}
  {\bibfnamefont {Y.}~\bibnamefont {Xiao}},\ }\href
  {https://doi.org/10.1038/s41586-020-2243-7} {\bibfield  {journal} {\bibinfo
  {journal} {Nature}\ }\textbf {\bibinfo {volume} {581}},\ \bibinfo {pages}
  {159} (\bibinfo {year} {2020})}\BibitemShut {NoStop}%
\bibitem [{\citenamefont {Kong}\ \emph {et~al.}(2020)\citenamefont {Kong},
  \citenamefont {Jim{\'e}nez-Mart{\'i}nez}, \citenamefont {Troullinou},
  \citenamefont {Lucivero}, \citenamefont {T{\'o}th},\ and\ \citenamefont
  {Mitchell}}]{Kong2020}%
  \BibitemOpen
  \bibfield  {author} {\bibinfo {author} {\bibfnamefont {J.}~\bibnamefont
  {Kong}}, \bibinfo {author} {\bibfnamefont {R.}~\bibnamefont
  {Jim{\'e}nez-Mart{\'i}nez}}, \bibinfo {author} {\bibfnamefont
  {C.}~\bibnamefont {Troullinou}}, \bibinfo {author} {\bibfnamefont {V.~G.}\
  \bibnamefont {Lucivero}}, \bibinfo {author} {\bibfnamefont {G.}~\bibnamefont
  {T{\'o}th}},\ and\ \bibinfo {author} {\bibfnamefont {M.~W.}\ \bibnamefont
  {Mitchell}},\ }\href {https://doi.org/10.1038/s41467-020-15899-1} {\bibfield
  {journal} {\bibinfo  {journal} {Nature Communications}\ }\textbf {\bibinfo
  {volume} {11}},\ \bibinfo {pages} {2415} (\bibinfo {year}
  {2020})}\BibitemShut {NoStop}%
\bibitem [{\citenamefont {Appel}\ \emph {et~al.}(2009)\citenamefont {Appel},
  \citenamefont {Windpassinger}, \citenamefont {Oblak}, \citenamefont {Hoff},
  \citenamefont {Kjaergaard},\ and\ \citenamefont {Polzik}}]{Appel2009}%
  \BibitemOpen
  \bibfield  {author} {\bibinfo {author} {\bibfnamefont {J.}~\bibnamefont
  {Appel}}, \bibinfo {author} {\bibfnamefont {P.~J.}\ \bibnamefont
  {Windpassinger}}, \bibinfo {author} {\bibfnamefont {D.}~\bibnamefont
  {Oblak}}, \bibinfo {author} {\bibfnamefont {U.~B.}\ \bibnamefont {Hoff}},
  \bibinfo {author} {\bibfnamefont {N.}~\bibnamefont {Kjaergaard}},\ and\
  \bibinfo {author} {\bibfnamefont {E.~S.}\ \bibnamefont {Polzik}},\ }\href
  {https://doi.org/10.1073/pnas.0901550106} {\bibfield  {journal} {\bibinfo
  {journal} {Proceedings of the National Academy of Sciences}\ }\textbf
  {\bibinfo {volume} {106}},\ \bibinfo {pages} {10960} (\bibinfo {year}
  {2009})},\ \Eprint {https://arxiv.org/abs/0810.3545} {arXiv:0810.3545}
  \BibitemShut {NoStop}%
\bibitem [{\citenamefont {Takano}\ \emph {et~al.}(2009)\citenamefont {Takano},
  \citenamefont {Fuyama}, \citenamefont {Namiki},\ and\ \citenamefont
  {Takahashi}}]{Takano2009}%
  \BibitemOpen
  \bibfield  {author} {\bibinfo {author} {\bibfnamefont {T.}~\bibnamefont
  {Takano}}, \bibinfo {author} {\bibfnamefont {M.}~\bibnamefont {Fuyama}},
  \bibinfo {author} {\bibfnamefont {R.}~\bibnamefont {Namiki}},\ and\ \bibinfo
  {author} {\bibfnamefont {Y.}~\bibnamefont {Takahashi}},\ }\href
  {https://doi.org/10.1103/PhysRevLett.102.033601} {\bibfield  {journal}
  {\bibinfo  {journal} {Phys. Rev. Lett.}\ }\textbf {\bibinfo {volume} {102}},\
  \bibinfo {pages} {033601} (\bibinfo {year} {2009})}\BibitemShut {NoStop}%
\bibitem [{\citenamefont {Sewell}\ \emph {et~al.}(2012)\citenamefont {Sewell},
  \citenamefont {Koschorreck}, \citenamefont {Napolitano}, \citenamefont
  {Dubost}, \citenamefont {Behbood},\ and\ \citenamefont
  {Mitchell}}]{Sewell2012}%
  \BibitemOpen
  \bibfield  {author} {\bibinfo {author} {\bibfnamefont {R.~J.}\ \bibnamefont
  {Sewell}}, \bibinfo {author} {\bibfnamefont {M.}~\bibnamefont {Koschorreck}},
  \bibinfo {author} {\bibfnamefont {M.}~\bibnamefont {Napolitano}}, \bibinfo
  {author} {\bibfnamefont {B.}~\bibnamefont {Dubost}}, \bibinfo {author}
  {\bibfnamefont {N.}~\bibnamefont {Behbood}},\ and\ \bibinfo {author}
  {\bibfnamefont {M.~W.}\ \bibnamefont {Mitchell}},\ }\href
  {https://doi.org/10.1103/PhysRevLett.109.253605} {\bibfield  {journal}
  {\bibinfo  {journal} {Physical Review Letters}\ }\textbf {\bibinfo {volume}
  {109}},\ \bibinfo {pages} {253605} (\bibinfo {year} {2012})}\BibitemShut
  {NoStop}%
\bibitem [{\citenamefont {Colangelo}\ \emph {et~al.}(2017)\citenamefont
  {Colangelo}, \citenamefont {Ciurana}, \citenamefont {Bianchet}, \citenamefont
  {Sewell},\ and\ \citenamefont {Mitchell}}]{Colangelo2017}%
  \BibitemOpen
  \bibfield  {author} {\bibinfo {author} {\bibfnamefont {G.}~\bibnamefont
  {Colangelo}}, \bibinfo {author} {\bibfnamefont {F.~M.}\ \bibnamefont
  {Ciurana}}, \bibinfo {author} {\bibfnamefont {L.~C.}\ \bibnamefont
  {Bianchet}}, \bibinfo {author} {\bibfnamefont {R.~J.}\ \bibnamefont
  {Sewell}},\ and\ \bibinfo {author} {\bibfnamefont {M.~W.}\ \bibnamefont
  {Mitchell}},\ }\href {https://doi.org/10.1038/nature21434} {\bibfield
  {journal} {\bibinfo  {journal} {Nature}\ }\textbf {\bibinfo {volume} {543}},\
  \bibinfo {pages} {525} (\bibinfo {year} {2017})}\BibitemShut {NoStop}%
\bibitem [{\citenamefont {Hemmer}\ \emph {et~al.}(2021)\citenamefont {Hemmer},
  \citenamefont {Monta\~no}, \citenamefont {Baragiola}, \citenamefont {Norris},
  \citenamefont {Shojaee}, \citenamefont {Deutsch},\ and\ \citenamefont
  {Jessen}}]{Hemmer2021}%
  \BibitemOpen
  \bibfield  {author} {\bibinfo {author} {\bibfnamefont {D.}~\bibnamefont
  {Hemmer}}, \bibinfo {author} {\bibfnamefont {E.}~\bibnamefont {Monta\~no}},
  \bibinfo {author} {\bibfnamefont {B.~Q.}\ \bibnamefont {Baragiola}}, \bibinfo
  {author} {\bibfnamefont {L.~M.}\ \bibnamefont {Norris}}, \bibinfo {author}
  {\bibfnamefont {E.}~\bibnamefont {Shojaee}}, \bibinfo {author} {\bibfnamefont
  {I.~H.}\ \bibnamefont {Deutsch}},\ and\ \bibinfo {author} {\bibfnamefont
  {P.~S.}\ \bibnamefont {Jessen}},\ }\href
  {https://doi.org/10.1103/PhysRevA.104.023710} {\bibfield  {journal} {\bibinfo
   {journal} {Phys. Rev. A}\ }\textbf {\bibinfo {volume} {104}},\ \bibinfo
  {pages} {023710} (\bibinfo {year} {2021})}\BibitemShut {NoStop}%
\bibitem [{\citenamefont {Est{\`e}ve}\ \emph {et~al.}(2008)\citenamefont
  {Est{\`e}ve}, \citenamefont {Gross}, \citenamefont {Weller}, \citenamefont
  {Giovanazzi},\ and\ \citenamefont {Oberthaler}}]{Esteve2008}%
  \BibitemOpen
  \bibfield  {author} {\bibinfo {author} {\bibfnamefont {J.}~\bibnamefont
  {Est{\`e}ve}}, \bibinfo {author} {\bibfnamefont {C.}~\bibnamefont {Gross}},
  \bibinfo {author} {\bibfnamefont {A.}~\bibnamefont {Weller}}, \bibinfo
  {author} {\bibfnamefont {S.}~\bibnamefont {Giovanazzi}},\ and\ \bibinfo
  {author} {\bibfnamefont {M.~K.}\ \bibnamefont {Oberthaler}},\ }\href
  {https://doi.org/10.1038/nature07332} {\bibfield  {journal} {\bibinfo
  {journal} {Nature}\ }\textbf {\bibinfo {volume} {455}},\ \bibinfo {pages}
  {1216} (\bibinfo {year} {2008})}\BibitemShut {NoStop}%
\bibitem [{\citenamefont {Gross}\ \emph {et~al.}(2010)\citenamefont {Gross},
  \citenamefont {Zibold}, \citenamefont {Nicklas}, \citenamefont {Est{\`e}ve},\
  and\ \citenamefont {Oberthaler}}]{Gross2010}%
  \BibitemOpen
  \bibfield  {author} {\bibinfo {author} {\bibfnamefont {C.}~\bibnamefont
  {Gross}}, \bibinfo {author} {\bibfnamefont {T.}~\bibnamefont {Zibold}},
  \bibinfo {author} {\bibfnamefont {E.}~\bibnamefont {Nicklas}}, \bibinfo
  {author} {\bibfnamefont {J.}~\bibnamefont {Est{\`e}ve}},\ and\ \bibinfo
  {author} {\bibfnamefont {M.~K.}\ \bibnamefont {Oberthaler}},\ }\href
  {https://doi.org/10.1038/nature08919} {\bibfield  {journal} {\bibinfo
  {journal} {Nature}\ }\textbf {\bibinfo {volume} {464}},\ \bibinfo {pages}
  {1165} (\bibinfo {year} {2010})}\BibitemShut {NoStop}%
\bibitem [{\citenamefont {Muessel}\ \emph {et~al.}(2014)\citenamefont
  {Muessel}, \citenamefont {Strobel}, \citenamefont {Linnemann}, \citenamefont
  {Hume},\ and\ \citenamefont {Oberthaler}}]{Muessel2014}%
  \BibitemOpen
  \bibfield  {author} {\bibinfo {author} {\bibfnamefont {W.}~\bibnamefont
  {Muessel}}, \bibinfo {author} {\bibfnamefont {H.}~\bibnamefont {Strobel}},
  \bibinfo {author} {\bibfnamefont {D.}~\bibnamefont {Linnemann}}, \bibinfo
  {author} {\bibfnamefont {D.~B.}\ \bibnamefont {Hume}},\ and\ \bibinfo
  {author} {\bibfnamefont {M.~K.}\ \bibnamefont {Oberthaler}},\ }\href
  {https://doi.org/10.1103/PhysRevLett.113.103004} {\bibfield  {journal}
  {\bibinfo  {journal} {Physical Review Letters}\ }\textbf {\bibinfo {volume}
  {113}},\ \bibinfo {pages} {103004} (\bibinfo {year} {2014})}\BibitemShut
  {NoStop}%
\bibitem [{\citenamefont {Stamper-Kurn}\ and\ \citenamefont
  {Ueda}(2013)}]{StamperKurn2013}%
  \BibitemOpen
  \bibfield  {author} {\bibinfo {author} {\bibfnamefont {D.~M.}\ \bibnamefont
  {Stamper-Kurn}}\ and\ \bibinfo {author} {\bibfnamefont {M.}~\bibnamefont
  {Ueda}},\ }\href {https://doi.org/10.1103/RevModPhys.85.1191} {\bibfield
  {journal} {\bibinfo  {journal} {Rev. Mod. Phys.}\ }\textbf {\bibinfo {volume}
  {85}},\ \bibinfo {pages} {1191} (\bibinfo {year} {2013})}\BibitemShut
  {NoStop}%
\bibitem [{\citenamefont {Kristensen}\ \emph {et~al.}(2019)\citenamefont
  {Kristensen}, \citenamefont {Christensen}, \citenamefont {Gajdacz},
  \citenamefont {Iglicki}, \citenamefont {Paw\l{}owski}, \citenamefont
  {Klempt}, \citenamefont {Sherson}, \citenamefont {Rzazewski}, \citenamefont
  {Hilliard},\ and\ \citenamefont {Arlt}}]{Kristensen2019}%
  \BibitemOpen
  \bibfield  {author} {\bibinfo {author} {\bibfnamefont {M.~A.}\ \bibnamefont
  {Kristensen}}, \bibinfo {author} {\bibfnamefont {M.~B.}\ \bibnamefont
  {Christensen}}, \bibinfo {author} {\bibfnamefont {M.}~\bibnamefont
  {Gajdacz}}, \bibinfo {author} {\bibfnamefont {M.}~\bibnamefont {Iglicki}},
  \bibinfo {author} {\bibfnamefont {K.}~\bibnamefont {Paw\l{}owski}}, \bibinfo
  {author} {\bibfnamefont {C.}~\bibnamefont {Klempt}}, \bibinfo {author}
  {\bibfnamefont {J.~F.}\ \bibnamefont {Sherson}}, \bibinfo {author}
  {\bibfnamefont {K.}~\bibnamefont {Rzazewski}}, \bibinfo {author}
  {\bibfnamefont {A.~J.}\ \bibnamefont {Hilliard}},\ and\ \bibinfo {author}
  {\bibfnamefont {J.~J.}\ \bibnamefont {Arlt}},\ }\href
  {https://doi.org/10.1103/PhysRevLett.122.163601} {\bibfield  {journal}
  {\bibinfo  {journal} {Phys. Rev. Lett.}\ }\textbf {\bibinfo {volume} {122}},\
  \bibinfo {pages} {163601} (\bibinfo {year} {2019})}\BibitemShut {NoStop}%
\bibitem [{\citenamefont {Politzer}(1996)}]{Politzer1996}%
  \BibitemOpen
  \bibfield  {author} {\bibinfo {author} {\bibfnamefont {H.~D.}\ \bibnamefont
  {Politzer}},\ }\href {https://doi.org/10.1103/PhysRevA.54.5048} {\bibfield
  {journal} {\bibinfo  {journal} {Phys. Rev. A}\ }\textbf {\bibinfo {volume}
  {54}},\ \bibinfo {pages} {5048} (\bibinfo {year} {1996})}\BibitemShut
  {NoStop}%
\bibitem [{\citenamefont {Navez}\ \emph {et~al.}(1997)\citenamefont {Navez},
  \citenamefont {Bitouk}, \citenamefont {Gajda}, \citenamefont {Idziaszek},\
  and\ \citenamefont {Rza\ifmmode \mbox{\c{}}\else
  \c{}\fi{}\ifmmode~\dot{z}\else \.{z}\fi{}ewski}}]{Navez1997}%
  \BibitemOpen
  \bibfield  {author} {\bibinfo {author} {\bibfnamefont {P.}~\bibnamefont
  {Navez}}, \bibinfo {author} {\bibfnamefont {D.}~\bibnamefont {Bitouk}},
  \bibinfo {author} {\bibfnamefont {M.}~\bibnamefont {Gajda}}, \bibinfo
  {author} {\bibfnamefont {Z.}~\bibnamefont {Idziaszek}},\ and\ \bibinfo
  {author} {\bibfnamefont {K.}~\bibnamefont {Rza\ifmmode \mbox{\c{}}\else
  \c{}\fi{}\ifmmode~\dot{z}\else \.{z}\fi{}ewski}},\ }\href
  {https://doi.org/10.1103/PhysRevLett.79.1789} {\bibfield  {journal} {\bibinfo
   {journal} {Phys. Rev. Lett.}\ }\textbf {\bibinfo {volume} {79}},\ \bibinfo
  {pages} {1789} (\bibinfo {year} {1997})}\BibitemShut {NoStop}%
\bibitem [{\citenamefont {Kuzmich}\ \emph {et~al.}(1998)\citenamefont
  {Kuzmich}, \citenamefont {Bigelow},\ and\ \citenamefont
  {Mandel}}]{Kuzmich1998}%
  \BibitemOpen
  \bibfield  {author} {\bibinfo {author} {\bibfnamefont {A.}~\bibnamefont
  {Kuzmich}}, \bibinfo {author} {\bibfnamefont {N.~P.}\ \bibnamefont
  {Bigelow}},\ and\ \bibinfo {author} {\bibfnamefont {L.}~\bibnamefont
  {Mandel}},\ }\href@noop {} {\bibfield  {journal} {\bibinfo  {journal}
  {Europhysical Letters}\ }\textbf {\bibinfo {volume} {42}},\ \bibinfo {pages}
  {481} (\bibinfo {year} {1998})}\BibitemShut {NoStop}%
\bibitem [{\citenamefont {Takahashi}\ \emph {et~al.}(1999)\citenamefont
  {Takahashi}, \citenamefont {Honda}, \citenamefont {Tanaka}, \citenamefont
  {Toyoda}, \citenamefont {Ishikawa},\ and\ \citenamefont
  {Yabuzaki}}]{Takahashi1999}%
  \BibitemOpen
  \bibfield  {author} {\bibinfo {author} {\bibfnamefont {Y.}~\bibnamefont
  {Takahashi}}, \bibinfo {author} {\bibfnamefont {K.}~\bibnamefont {Honda}},
  \bibinfo {author} {\bibfnamefont {N.}~\bibnamefont {Tanaka}}, \bibinfo
  {author} {\bibfnamefont {K.}~\bibnamefont {Toyoda}}, \bibinfo {author}
  {\bibfnamefont {K.}~\bibnamefont {Ishikawa}},\ and\ \bibinfo {author}
  {\bibfnamefont {T.}~\bibnamefont {Yabuzaki}},\ }\href
  {https://doi.org/10.1103/PhysRevA.60.4974} {\bibfield  {journal} {\bibinfo
  {journal} {Physical Review A}\ }\textbf {\bibinfo {volume} {60}},\ \bibinfo
  {pages} {4974} (\bibinfo {year} {1999})}\BibitemShut {NoStop}%
\bibitem [{\citenamefont {Hammerer}\ \emph {et~al.}(2004)\citenamefont
  {Hammerer}, \citenamefont {M\o{}lmer}, \citenamefont {Polzik},\ and\
  \citenamefont {Cirac}}]{Hammerer2004}%
  \BibitemOpen
  \bibfield  {author} {\bibinfo {author} {\bibfnamefont {K.}~\bibnamefont
  {Hammerer}}, \bibinfo {author} {\bibfnamefont {K.}~\bibnamefont {M\o{}lmer}},
  \bibinfo {author} {\bibfnamefont {E.~S.}\ \bibnamefont {Polzik}},\ and\
  \bibinfo {author} {\bibfnamefont {J.~I.}\ \bibnamefont {Cirac}},\ }\href
  {https://doi.org/10.1103/PhysRevA.70.044304} {\bibfield  {journal} {\bibinfo
  {journal} {Phys. Rev. A}\ }\textbf {\bibinfo {volume} {70}},\ \bibinfo
  {pages} {044304} (\bibinfo {year} {2004})}\BibitemShut {NoStop}%
\bibitem [{\citenamefont {Baragiola}\ \emph {et~al.}(2014)\citenamefont
  {Baragiola}, \citenamefont {Norris}, \citenamefont {Monta{\~{n}}o},
  \citenamefont {Mickelson}, \citenamefont {Jessen},\ and\ \citenamefont
  {Deutsch}}]{Baragiola2014}%
  \BibitemOpen
  \bibfield  {author} {\bibinfo {author} {\bibfnamefont {B.~Q.}\ \bibnamefont
  {Baragiola}}, \bibinfo {author} {\bibfnamefont {L.~M.}\ \bibnamefont
  {Norris}}, \bibinfo {author} {\bibfnamefont {E.}~\bibnamefont
  {Monta{\~{n}}o}}, \bibinfo {author} {\bibfnamefont {P.~G.}\ \bibnamefont
  {Mickelson}}, \bibinfo {author} {\bibfnamefont {P.~S.}\ \bibnamefont
  {Jessen}},\ and\ \bibinfo {author} {\bibfnamefont {I.~H.}\ \bibnamefont
  {Deutsch}},\ }\href {https://doi.org/10.1103/PhysRevA.89.033850} {\bibfield
  {journal} {\bibinfo  {journal} {Physical Review A}\ }\textbf {\bibinfo
  {volume} {89}},\ \bibinfo {pages} {033850} (\bibinfo {year}
  {2014})}\BibitemShut {NoStop}%
\bibitem [{\citenamefont {Mitchell}\ and\ \citenamefont
  {Palacios~Alvarez}(2020)}]{Mitchell2020}%
  \BibitemOpen
  \bibfield  {author} {\bibinfo {author} {\bibfnamefont {M.~W.}\ \bibnamefont
  {Mitchell}}\ and\ \bibinfo {author} {\bibfnamefont {S.}~\bibnamefont
  {Palacios~Alvarez}},\ }\href {https://doi.org/10.1103/RevModPhys.92.021001}
  {\bibfield  {journal} {\bibinfo  {journal} {Rev. Mod. Phys.}\ }\textbf
  {\bibinfo {volume} {92}},\ \bibinfo {pages} {021001} (\bibinfo {year}
  {2020})}\BibitemShut {NoStop}%
\bibitem [{\citenamefont {Windpassinger}\ \emph {et~al.}(2009)\citenamefont
  {Windpassinger}, \citenamefont {Kubasik}, \citenamefont {Koschorreck},
  \citenamefont {Boisen}, \citenamefont {Kj{\ae}rgaard}, \citenamefont
  {Polzik},\ and\ \citenamefont {M\"{u}ller}}]{Windpassinger2009}%
  \BibitemOpen
  \bibfield  {author} {\bibinfo {author} {\bibfnamefont {P.~J.}\ \bibnamefont
  {Windpassinger}}, \bibinfo {author} {\bibfnamefont {M.}~\bibnamefont
  {Kubasik}}, \bibinfo {author} {\bibfnamefont {M.}~\bibnamefont
  {Koschorreck}}, \bibinfo {author} {\bibfnamefont {A.}~\bibnamefont {Boisen}},
  \bibinfo {author} {\bibfnamefont {N.}~\bibnamefont {Kj{\ae}rgaard}}, \bibinfo
  {author} {\bibfnamefont {E.~S.}\ \bibnamefont {Polzik}},\ and\ \bibinfo
  {author} {\bibfnamefont {J.~H.}\ \bibnamefont {M\"{u}ller}},\ }\href
  {https://doi.org/10.1088/0957-0233/20/5/055301} {\bibfield  {journal}
  {\bibinfo  {journal} {Measurement Science and Technology}\ }\textbf {\bibinfo
  {volume} {20}},\ \bibinfo {pages} {055301} (\bibinfo {year}
  {2009})}\BibitemShut {NoStop}%
\bibitem [{\citenamefont {Ciurana}\ \emph {et~al.}(2016)\citenamefont
  {Ciurana}, \citenamefont {Colangelo}, \citenamefont {Sewell},\ and\
  \citenamefont {Mitchell}}]{Ciurana2016}%
  \BibitemOpen
  \bibfield  {author} {\bibinfo {author} {\bibfnamefont {F.~M.}\ \bibnamefont
  {Ciurana}}, \bibinfo {author} {\bibfnamefont {G.}~\bibnamefont {Colangelo}},
  \bibinfo {author} {\bibfnamefont {R.~J.}\ \bibnamefont {Sewell}},\ and\
  \bibinfo {author} {\bibfnamefont {M.~W.}\ \bibnamefont {Mitchell}},\ }\href
  {https://doi.org/10.1364/OL.41.002946} {\bibfield  {journal} {\bibinfo
  {journal} {Opt. Lett.}\ }\textbf {\bibinfo {volume} {41}},\ \bibinfo {pages}
  {2946} (\bibinfo {year} {2016})}\BibitemShut {NoStop}%
\bibitem [{\citenamefont {Takai}\ \emph {et~al.}(2023)\citenamefont {Takai},
  \citenamefont {Shibata}, \citenamefont {Sekiguchi},\ and\ \citenamefont
  {Hirano}}]{Takai2023}%
  \BibitemOpen
  \bibfield  {author} {\bibinfo {author} {\bibfnamefont {J.}~\bibnamefont
  {Takai}}, \bibinfo {author} {\bibfnamefont {K.}~\bibnamefont {Shibata}},
  \bibinfo {author} {\bibfnamefont {N.}~\bibnamefont {Sekiguchi}},\ and\
  \bibinfo {author} {\bibfnamefont {T.}~\bibnamefont {Hirano}},\ }\href
  {https://doi.org/10.1103/PhysRevA.107.053308} {\bibfield  {journal} {\bibinfo
   {journal} {Phys. Rev. A}\ }\textbf {\bibinfo {volume} {107}},\ \bibinfo
  {pages} {053308} (\bibinfo {year} {2023})}\BibitemShut {NoStop}%
\bibitem [{\citenamefont {Deutsch}\ and\ \citenamefont
  {Jessen}(2010)}]{Deutsch2010}%
  \BibitemOpen
  \bibfield  {author} {\bibinfo {author} {\bibfnamefont {I.~H.}\ \bibnamefont
  {Deutsch}}\ and\ \bibinfo {author} {\bibfnamefont {P.~S.}\ \bibnamefont
  {Jessen}},\ }\href {https://doi.org/10.1016/j.optcom.2009.10.059} {\bibfield
  {journal} {\bibinfo  {journal} {Optics Communications}\ }\textbf {\bibinfo
  {volume} {283}},\ \bibinfo {pages} {681} (\bibinfo {year}
  {2010})}\BibitemShut {NoStop}%
\bibitem [{\citenamefont {Steffen}\ \emph {et~al.}(2013)\citenamefont
  {Steffen}, \citenamefont {Alt}, \citenamefont {Genske}, \citenamefont
  {Meschede}, \citenamefont {Robens},\ and\ \citenamefont
  {Alberti}}]{Steffen2013}%
  \BibitemOpen
  \bibfield  {author} {\bibinfo {author} {\bibfnamefont {A.}~\bibnamefont
  {Steffen}}, \bibinfo {author} {\bibfnamefont {W.}~\bibnamefont {Alt}},
  \bibinfo {author} {\bibfnamefont {M.}~\bibnamefont {Genske}}, \bibinfo
  {author} {\bibfnamefont {D.}~\bibnamefont {Meschede}}, \bibinfo {author}
  {\bibfnamefont {C.}~\bibnamefont {Robens}},\ and\ \bibinfo {author}
  {\bibfnamefont {A.}~\bibnamefont {Alberti}},\ }\href
  {https://doi.org/10.1063/1.4847075} {\bibfield  {journal} {\bibinfo
  {journal} {Review of Scientific Instruments}\ }\textbf {\bibinfo {volume}
  {84}},\ \bibinfo {pages} {126103} (\bibinfo {year} {2013})}\BibitemShut
  {NoStop}%
\bibitem [{\citenamefont {Wood}\ \emph {et~al.}(2016)\citenamefont {Wood},
  \citenamefont {Turner},\ and\ \citenamefont {Anderson}}]{Wood2016}%
  \BibitemOpen
  \bibfield  {author} {\bibinfo {author} {\bibfnamefont {A.~A.}\ \bibnamefont
  {Wood}}, \bibinfo {author} {\bibfnamefont {L.~D.}\ \bibnamefont {Turner}},\
  and\ \bibinfo {author} {\bibfnamefont {R.~P.}\ \bibnamefont {Anderson}},\
  }\href {https://doi.org/10.1103/PhysRevA.94.052503} {\bibfield  {journal}
  {\bibinfo  {journal} {Phys. Rev. A}\ }\textbf {\bibinfo {volume} {94}},\
  \bibinfo {pages} {052503} (\bibinfo {year} {2016})}\BibitemShut {NoStop}%
\bibitem [{\citenamefont {Shibata}\ \emph {et~al.}(2021)\citenamefont
  {Shibata}, \citenamefont {Sekiguchi},\ and\ \citenamefont
  {Hirano}}]{Shibata2021}%
  \BibitemOpen
  \bibfield  {author} {\bibinfo {author} {\bibfnamefont {K.}~\bibnamefont
  {Shibata}}, \bibinfo {author} {\bibfnamefont {N.}~\bibnamefont {Sekiguchi}},\
  and\ \bibinfo {author} {\bibfnamefont {T.}~\bibnamefont {Hirano}},\ }\href
  {https://doi.org/10.1103/PhysRevA.103.043335} {\bibfield  {journal} {\bibinfo
   {journal} {Phys. Rev. A}\ }\textbf {\bibinfo {volume} {103}},\ \bibinfo
  {pages} {043335} (\bibinfo {year} {2021})}\BibitemShut {NoStop}%
\bibitem [{\citenamefont {Smith}\ \emph {et~al.}(2004)\citenamefont {Smith},
  \citenamefont {Chaudhury}, \citenamefont {Silberfarb}, \citenamefont
  {Deutsch},\ and\ \citenamefont {Jessen}}]{Smith2004}%
  \BibitemOpen
  \bibfield  {author} {\bibinfo {author} {\bibfnamefont {G.~A.}\ \bibnamefont
  {Smith}}, \bibinfo {author} {\bibfnamefont {S.}~\bibnamefont {Chaudhury}},
  \bibinfo {author} {\bibfnamefont {A.}~\bibnamefont {Silberfarb}}, \bibinfo
  {author} {\bibfnamefont {I.~H.}\ \bibnamefont {Deutsch}},\ and\ \bibinfo
  {author} {\bibfnamefont {P.~S.}\ \bibnamefont {Jessen}},\ }\href
  {https://doi.org/10.1103/PhysRevLett.93.163602} {\bibfield  {journal}
  {\bibinfo  {journal} {Phys. Rev. Lett.}\ }\textbf {\bibinfo {volume} {93}},\
  \bibinfo {pages} {163602} (\bibinfo {year} {2004})}\BibitemShut {NoStop}%
\bibitem [{\citenamefont {Koschorreck}\ \emph {et~al.}(2010)\citenamefont
  {Koschorreck}, \citenamefont {Napolitano}, \citenamefont {Dubost},\ and\
  \citenamefont {Mitchell}}]{Koschorreck2010}%
  \BibitemOpen
  \bibfield  {author} {\bibinfo {author} {\bibfnamefont {M.}~\bibnamefont
  {Koschorreck}}, \bibinfo {author} {\bibfnamefont {M.}~\bibnamefont
  {Napolitano}}, \bibinfo {author} {\bibfnamefont {B.}~\bibnamefont {Dubost}},\
  and\ \bibinfo {author} {\bibfnamefont {M.~W.}\ \bibnamefont {Mitchell}},\
  }\href {https://doi.org/10.1103/PhysRevLett.105.093602} {\bibfield  {journal}
  {\bibinfo  {journal} {Phys. Rev. Lett.}\ }\textbf {\bibinfo {volume} {105}},\
  \bibinfo {pages} {093602} (\bibinfo {year} {2010})}\BibitemShut {NoStop}%
\bibitem [{\citenamefont {Burnett}\ \emph {et~al.}(1996)\citenamefont
  {Burnett}, \citenamefont {Julienne},\ and\ \citenamefont
  {Suominen}}]{Burnett1996}%
  \BibitemOpen
  \bibfield  {author} {\bibinfo {author} {\bibfnamefont {K.}~\bibnamefont
  {Burnett}}, \bibinfo {author} {\bibfnamefont {P.~S.}\ \bibnamefont
  {Julienne}},\ and\ \bibinfo {author} {\bibfnamefont {K.-A.}\ \bibnamefont
  {Suominen}},\ }\href {https://doi.org/10.1103/PhysRevLett.77.1416} {\bibfield
   {journal} {\bibinfo  {journal} {Phys. Rev. Lett.}\ }\textbf {\bibinfo
  {volume} {77}},\ \bibinfo {pages} {1416} (\bibinfo {year}
  {1996})}\BibitemShut {NoStop}%
\end{thebibliography}
%

\end{document}